\documentclass[review,12pt]{elsarticle}

\usepackage{bm}  
\usepackage{amssymb}
\usepackage{amsthm}
\usepackage{amsmath}
\usepackage[linesnumbered,ruled,lined]{algorithm2e}
\graphicspath{{figs/}} 
\usepackage{booktabs}
\usepackage{multirow}
\usepackage{hyperref}
\newcommand{\tr}{\mathop{\mathrm{tr}}}

\usepackage[outline]{contour}
\newcommand*{\fancy}[1]{{\color{white}\contour{black}{#1}}}
\usepackage{booktabs}
\usepackage{caption} 
\journal{X.X.X.}

\begin{document}

\begin{frontmatter}



\title{A unified SPH framework for shell-related interactions}
\author[myfirstaddress]{Dong Wu}
\ead{dong.wu@tum.de}
\author[mysecondaddress]{Shuaihao Zhang}
\ead{shzhangce@ust.hk}
\author[mythirdaddress]{Weiyi Kong}
\ead{kong@virtonomy.io}
\author[myfirstaddress]{Xiangyu Hu\corref{mycorrespondingauthor}}
\cortext[mycorrespondingauthor]{Corresponding author.}
\ead{xiangyu.hu@tum.de}

\address[myfirstaddress]{Chair of Aerodynamics and Fluid Mechanics, Technical University of Munich, 85748 Garching, Germany}
\address[mysecondaddress]{Department of Civil and Environmental Engineering, Hong Kong University of Science and Technology, Clearwater Bay, Kowloon, Hong Kong SAR, China}
\address[mythirdaddress]{Virtonomy GmbH, 80336 Munich, Germany}

\begin{abstract}
A unified Smoothed Particle Hydrodynamics (SPH) framework 
is proposed to simulate interaction dynamics 
involving thin shells modeled by a reduced-dimensional, 
single-layer particle discretization, 
as opposed to full-dimensional SPH solids.
The framework encompasses one-sided fluid–shell interactions, 
with the fluid present on only one side of the shell, 
as well as solid–shell, shell–shell, and shell-self interactions
The study introduces a novel concept of imaginary shell contact particles, 
generated by projecting real shell particles along the local normal direction 
within the cut-off radius of the fluid particle,
thereby mapping this reduced-dimensional shell model 
into a full-dimensional representation.
With the volume of the imaginary particles 
defined based on the local shell curvature, 
the projection preserves kernel completeness for fluid-shell interactions 
while leaving the fluid-structure interaction (FSI) dynamics unchanged, 
such that the fluid-shell coupling algorithm 
is the same as in standard fluid-solid coupling.
In addition,
a particle-to-particle contact model for solid-solid interactions is developed 
by analogy to fluid dynamics: 
a contact density is computed using a fluid-style density initialization, 
and the resulting contact forces 
follow a momentum-equation-inspired formulation. 
Combined with the projection strategy, 
this contact formulation is directly extended 
to efficiently handle shell-related contact problems. 
The proposed method is validated using a series of benchmark tests, 
demonstrating stable and accurate performance
across diverse interaction scenarios.

\end{abstract}
\begin{keyword}
Fluid-shell interaction  \sep Solid-shell contact \sep Shell-shell contact \sep Shell-self contact \sep SPH.
\end{keyword}

\end{frontmatter}


\section{Introduction}\label{sec:introduction}
Smoothed Particle Hydrodynamics (SPH), 
a fully Lagrangian and mesh-free method, 
has attracted increasing attention over recent decades 
\cite{randles1996smoothed, luo2021particle, zhang2022review, xu2023methodology, khayyer2023preface}. 
In SPH, 
continua are discretized into particles, 
and the governing equations solved via particle interactions 
mediated by a Gaussian-like kernel function 
\cite{monaghan2005smoothed, liu2010smoothed, monaghan2012smoothed}. 
Owing to its particle-interaction-based formulation, 
which aligns naturally with abstractions 
underlying a wide range of physical systems, 
SPH has proven particularly effective for multi-physics simulations 
within a unified computational framework 
\cite{zhang2021sphinxsys, sun2021accurate}.

Fluid–structure interaction (FSI), 
particularly involving flexible and deformable elastic structures,
represents a typical multi-physical system 
in which fluid and structural dynamics are tightly coupled. 
Due to the intrinsic complexity of the surrounding flow 
and the potentially large deformation of the structure, 
computational studies of FSI problems remain highly challenging. 
Accordingly, 
much of the SPH literature has focused on fluid–structure coupling
\cite{antoci2007numerical, rafiee2009sph, han2018sph, khayyer2018enhanced, liu2019smoothed, sun2019study, zhang2019smoothed, pearl2021sph, khayyer2024improved}. 
In these models,
pressure and viscous forces across the fluid–structure interface are typically evaluated 
\cite{adami2012generalized, han2018sph, zhang2021multi}, 
and density initialization schemes are often adapted 
in the presence of structure boundaries to alleviate kernel truncation effects 
\cite{colagrossi2003numerical, zhang2017generalized, zhang2020dual}.

In many practical FSI scenarios, 
structural dynamics also involve contact between interacting bodies or self-contact.
Existing SPH contact methods can generally be categorized into 
particle-to-particle and particle-to-segment approaches \cite{zhan2020sph}. 
In particle-to-particle contact, 
particles are idealized as rigid spheres 
and penalty forces are introduced to prevent interpenetration 
\cite{combescure2008modelling, seo2008application, islam2019total}. 
In particle-to-segment contact, 
normal penalty forces are defined 
with respect to a segment-based surface representation, 
which facilitates the incorporation of friction 
\cite{gutfraind1997smoothed, bui2014novel, dong2016smoothed}.

Despite the progress in SPH-based FSI and structural contact modeling, 
interactions involving shells, 
which are thin structures modeled by reduced-dimensional, 
single-layer particle discretizations, 
remain comparatively less explored compared with 
those for full-dimensional SPH solids.
Owing to their thin-walled nature and reduced-dimensional representation, 
shells pose additional challenges for both fluid coupling and contact modeling. 
Only a limited number of studies have demonstrated the capability of SPH 
to capture deformation of shell structures under fluid loading, 
as well as the feedback of shell motion on the surrounding flow 
\cite{combescure2008modelling, maurel2009full, peng2021coupling, li2022algorithm, gao2024three, bao2024entirely}. 
Two strategies are most commonly adopted. 
One is the normal-flux method \cite{leroy2014unified}, 
which distinguishes neighboring fluid particles located on opposite sides of a shell 
and has mainly been used for two-sided FSI, 
where the shell is wetted by fluid on both sides 
\cite{peng2021coupling, gao2024three, bao2024entirely}.
The other is a particle-to-particle contact-based treatment, 
which has been employed primarily for one-sided FSI 
and has also been used in example settings involving solid–shell contact 
\cite{combescure2008modelling, maurel2009full}. 

Nevertheless, 
a unified treatment covering fluid–shell coupling together with solid–shell, shell–shell, and shell-self contact is still lacking. 
In this paper, 
we propose a unified SPH framework for shell-related interaction dynamics. 
The framework introduces imaginary shell contact particles 
by projecting real shell particles along the local normal direction 
within a fluid particle’s cut-off radius, 
thereby mapping the reduced-dimensional shell model 
into an effective full-dimensional representation. 
The volume of these imaginary particles 
is determined from the local shell curvature, 
which preserves kernel completeness for fluid particles 
near the shell boundary while leaving the underlying FSI dynamics unchanged, 
so that fluid–shell coupling follows the same algorithm 
as standard fluid–solid coupling. 
In addition, 
a particle-to-particle contact model for solid–solid interactions 
is developed by analogy to fluid dynamics, 
where a contact density is computed using a fluid-style density initialization 
and the corresponding contact forces 
follow a momentum-equation-inspired formulation. 
Combined with the projection strategy, 
this contact formulation is directly extended 
to efficiently handle shell-related contact problems.
The proposed framework is validated through a series of benchmark tests, 
demonstrating stability, accuracy, and versatility 
across diverse interaction scenarios. 
In addition, 
a complex industrial case study, 
a half-filled oil tank impacted by a struck, 
is simulated to further showcase the applicability of the proposed formulation, 
involving both FSI and contact events.

The structure of this paper is as follows. 
Section \ref{sec:governing_eq} describes 
the governing equations for fluid, solid, and shell models.
The interaction models 
including preliminary working of the fluid-solid interaction framework
and the present shell-related interaction models
are detailed in Section \ref{sec:interaction}. 
Numerical examples are presented and discussed in Section \ref{sec:examples}.
Section~\ref{sec:conclusion} concludes the paper with a summary and outlook.
To support future research and reproducibility,
all computational codes utilized in this study 
are publicly available through the SPHinXsys \cite{zhang2021sphinxsys, zhang2020sphinxsys} project website 
 at \url{https://www.sphinxsys.org}.

\section{Governing equations} \label{sec:governing_eq}
Before moving to detailed description of the interaction modeling, 
we first introduce the governing equations for fluid, solid and shell dynamics.
\subsection{Fluid dynamics equations}
The governing equations in updated Lagrangian formulation for an isothermal and Newtonian fluid flow  are the mass conservation equation
\begin{equation}\label{mass-conservation-f}
	\frac{\text{d} \rho}{\text{d} t}  = -\rho \nabla \cdot\mathbf{v} ,
\end{equation}
and the momentum conservation equation
\begin{equation}\label{momentum-conservation-f}
	\rho \frac{\text{d} \mathbf{v}}{\text{d} t}  = -\nabla p + \eta \nabla^{2} \mathbf{v},
\end{equation}
where $\rho$ is the density, 
$t$ the time,
$\mathbf{v}$ the velocity,
$p$ the pressure, 
$\eta$ the dynamics viscosity, 
and $\text{d}(\bullet)/\text{d}t=\partial(\bullet)
/\partial t+\boldsymbol{\rm v}\cdot\nabla(\bullet)$ refers 
to the material derivative.
An artificial equation of state (EoS) for weakly compressible 
flows is used to close the system of 
Eqs. \eqref{mass-conservation-f} and \eqref{momentum-conservation-f} as 
\begin{equation}\label{eq:eos}
	p= {c^F}^2\left(\rho-\rho_{0}\right).
\end{equation}
Here, $\rho_{0}$ is the initial density, and $c^F$ denotes 
the artificial sound speed. 
Setting $c^F = 10 U_{max}$, where $U_{max}$ represents the 
anticipated maximum fluid speed, fulfills the weakly compressible 
assumption where the density variation remains around $1\%$ 
\cite{morris1997modeling}.
\subsection{Solid dynamics equations}
Following the convention in continuum mechanics, 
the initial position $\mathbf{r}^0$ of a material point is defined in the initial reference configuration, 
and the current position $\mathbf{r}$ in the deformed current configuration. 
Then the displacement $\mathbf{u}$ of a material point, 
which is obtained by the difference between its current position and initial reference position, 
can be obtained as
\begin{equation}\label{trajectory}
	\mathbf{u} =  \mathbf{r} - \mathbf{r}^0 .
\end{equation}
Thus, the deformation tensor, is defined by 
\begin{equation}\label{deformation-tensor}
	\mathbb{F}  = \nabla^{0} \mathbf{u}  + \mathbb{I} ,
\end{equation}
where $\mathbb{I}$ denotes the identity matrix, 
and $\nabla^{0} \equiv \frac{\partial}{\partial \mathbf{r}^0}$ stands for the gradient operator with respect to the initial reference configuration. 
In the total Lagrangian formulation, the mass conservation equation is given by
\begin{equation}\label{mass-conservation-s}
	\rho  -  \rho^0 J ^{-1} = 0 ,
\end{equation}
where $\rho^0$ is the initial reference density and $J = \det(\mathbb{F})$ the Jacobian determinant of deformation tensor $\mathbb{F}$. 
The momentum conservation equation reads
\begin{equation}\label{momentum-conservation-s}
	\rho^0 \frac{d \mathbf{v}}{d t}  =  \nabla^{0} \cdot \mathbb{P},
\end{equation}
where $\mathbb{P}$ denotes the first Piola-Kirchhoff stress tensor, 
which relates the forces in the current configuration to the areas in the initial reference configuration.
For an ideal elastic or Kirchhoff material, $\mathbb{P}$ is given by
\begin{equation}\label{linear-elasticity}
	\mathbb{P} = \mathbb{F} \mathbb{S}, 
\end{equation}
where $\mathbb{S}$ represents the second Piola-Kirchhoff stress which is evaluated via the constitutive equation relating $\mathbb{F}$ with 
the Green-Lagrangian strain tensor $\mathbb{E}$, defined as 
\begin{equation}\label{Lagrangian-strain}
	\mathbb{E} = \frac{1}{2} \left( \mathbb{F}^{T}\mathbb{F} - \mathbb{I}\right) .
\end{equation}

In particular, when the material is linear elastic and isotropic, the constitutive equation is simply given by
\begin{eqnarray}\label{isotropic-linear-elasticity}
	\mathbb{S} & = & K \tr\left(\mathbb{E}\right)  \mathbb{I} + 2 G \left(\mathbb{E} - \frac{1}{3}\tr\left(\mathbb{E}\right)  \mathbb{I} \right) \nonumber \\
	& = & \lambda \tr\left(\mathbb{E}\right) \mathbb{I} + 2 \mu \mathbb{E} ,
\end{eqnarray}
where $\lambda$ and $\mu$ are Lem$\acute{e}$ parameters, 
$K = \lambda + (2\mu/3)$ the bulk modulus and $G = \mu$ the shear modulus. 
The relation between the two modulus is given by
\begin{equation}\label{relation-modulus}
	E = 2G \left(1+2\nu\right) = 3K\left(1 - 2\nu\right)
\end{equation}
with $E$ denotes the Young's modulus and $\nu$ the Poisson ratio. 
Note that the sound speed of solid structure is defined as $c^{S} = \sqrt{\frac{K}{\rho}}$. 

\subsection{Shell dynamics equations}
The Uflyand–Mindlin plate theory \cite{uflyand1948wave, mindlin1951influence} 
is considered to account for transverse shear stress 
which is significant for moderately thick plates and shells.
The theory implies that the thin-structure behavior can be represented 
by one layer of material points at its mid-surface.

We introduce $\bm{X}$ to represent the global coordinate system, 
and $\bm{\xi}$ and $\bm{x}$, 
associated with so-called pseudo-normal vector $ \bm{n}$ of shell, 
to denote the initial and current local coordinate systems, respectively.
The kinematics of shell can be constructed in the local coordinates 
denoted with the superscript $\left( \bullet \right)^L$.
Each material point possesses degrees of freedom: 
the translation $ \bm{u}^L $ 
and rotation $\bm{\theta}^L$.
The pseudo-normal vector is also presented in initial local coordinates by 
$ \bm{n}^L$.
The local position $ \bm{r}^L $ of a material point can be expressed as 
\begin{equation}
	\bm{r}^L
	= \bm{r}^L_m
	+ \chi \bm{n}^L,  
	\quad \chi  \in \left[- d^0/2,d^0/2 \right], 
\end{equation}
where $d^0$ is the shell thickness, 
$ \bm{r}_m $ the position of the material point at the mid-surface 
with the subscript $\left( \bullet \right)_m$ denoting the mid-surface. 
Note that the shell thickness is assumed to be unchanged.
The local displacement $ \bm{u}^L$ can thus be obtained by 
\begin{equation}
	\bm{u}^L 
	= \bm{u}^L_m + \chi \Delta \bm{n}^L,
\end{equation}
where $\Delta \bm{n}^L = \bm{n}^L - \bm{n}^{0, L} $.
The local deformation gradient tensor of 3D shells can be defined as
\begin{equation}\label{shell_deformation_tensor}
	\mathbb{F}^L  = 
	\mathbb{F}_m^L + \chi \mathbb{F}_n^L
	=\nabla^{0, L} \bm{r}^L  
	+ \chi \left(\nabla^{0, L} \bm{n}^L
	- \nabla^{0, L} \bm{n}^{0, L}\right),
\end{equation}
where $\nabla^{0, L} \equiv  \partial / \partial \bm{\xi}$ 
is the gradient operators defined in the initial local configuration.

With the deformation gradient tensor $\mathbb{F}^L$, 
the Eulerian Almansi strain $\fancy{$\epsilon$}^l$ 
in current local coordinates
can be obtained as 
\begin{equation}\label{Almansi_strain}
	\fancy{$\epsilon$}^l 
	= \frac{1}{2} \mathbb{Q} \left(\mathbb{Q}^0\right)^{\operatorname{T}}
	\left(\mathbb{I} - \left(\mathbb{F}^L\right)^{-\operatorname{T}} 
	\left(\mathbb{F}^L\right)^{-1}\right)
	\mathbb{Q}^0 \mathbb{Q}^{\operatorname{T}}, 
\end{equation}
where $\mathbb{Q}^0$ is
the orthogonal transformation matrix from the global to initial local coordinates, 
and $\mathbb{Q}$ 
the transformation matrix from the global to current local coordinates.
Strain in the current local coordinates is required for strain and stress correction;
further details are available in our previous work \cite{wu2024sph}.
When the material is linear and isotropic, 
the Cauchy stress $\fancy{$\sigma$}^l$ reads 
\begin{equation}\label{constitutive_relation}
	\begin{split}
		\fancy{$\sigma$}^l = \lambda\operatorname{tr}\left( \fancy{$\epsilon$}^l \right) \mathbb{I} + 2\mu \fancy{$\epsilon$}^l. \\ 
	\end{split}
\end{equation}

The conservation equations of shell are 
\begin{equation} \label{plate-conservation-equation}
	\begin{cases}
		d^0 \rho ^0 \frac{\operatorname{d} \mathbf{v}_m}{\operatorname{d} t}  
		= J_m \left(\mathbb{F}_m\right)^{\operatorname{-T}} \nabla^0 
		\cdot \mathbb{N}^{\operatorname{T}} \\
		\frac{(d^0)^3}{12}\rho^0 \frac{\operatorname{d}^2 \mathbf{n}}{\operatorname{d} t^2}  
		= J_m \left(\mathbb{F}_m\right)^{\operatorname{-T}} \nabla^0  
		\cdot \mathbb{M}^{\operatorname{T}}
		+ J_m \mathbb{Q}^{\operatorname{T}}{\bf{Q}}^l, 
	\end{cases}
\end{equation}
where $\mathbb{N} = \mathbb{Q}^{\operatorname{T}} \mathbb{N}^l \mathbb{Q}$ 
and $\mathbb{M} = \mathbb{Q}^{\operatorname{T}} \mathbb{M}^l \mathbb{Q}$ 
are the stress and moment resultants, respectively, in global coordinates, 
${\bf{Q}}^l$ is the shear stress resultant, 
and $J_m = \det(\mathbb{F}_m)$.
Here, $\mathbb{N}^l$, $\mathbb{M}^l$ and ${\bf{Q}}^l$ is the integration 
of the Cauchy stress $\fancy{$\sigma$}^l$ along the shell thickness, 
and more details are referred to our previous work \cite{wu2024sph}.

\section{Interaction modeling} \label{sec:interaction}
Before moving to detailed description of the shell-related interaction method, 
we first briefly summarize our previous modeling of fluid-solid interaction 
with unified SPH formulation 
and more details are referred to Refs. \cite{han2018sph, zhang2021multi}. 
\subsection{Preliminary work}
\subsubsection{SPH discretization}
A SPH discretization of the mass and momentum conservation equations, 
Eqs. \eqref{mass-conservation-f}  and 
\eqref{momentum-conservation-f}, 
can be respectively derived as
\begin{equation} \label{eq:mass_equation}
	\frac{\text{d} \rho_i}{\text{d} t}  = 2\rho_i \sum_{j}  V_j (\mathbf{v}_i - \widetilde{\mathbf{v}}_{ij}) \nabla_i W_{ij}^{h^F},
\end{equation}
and
\begin{equation}\label{fluid-model}
	m_i \frac{\text{d} \mathbf{v}_i}{\text{d} t}  
	=  - 2  \sum_j V_i V_j \widetilde{p}_{ij} \nabla_i W_{ij}^{h^F} + 
	2\sum_j \eta V_i V_j \frac{{{\mathbf{v}}_{ij}}}{{{r}_{ij}}} \frac{\partial W_{ij}^{h^F}}{\partial {{r}_{ij}}} + \mathbf{g} +
	\mathbf{f}_i^{S:p}\left(h^F\right) + \mathbf{f}_i^{S:v}\left(h^F\right),
\end{equation}
where $V_i$ denotes the particle volume, 
$m_i$ the particle mass, 
$\mathbf{g}$ the body force, and 
$\nabla_i W_{ij}^{h^F} = \nabla_i W(\mathbf{r}_{ij}, h^F) =  \frac{\partial W_{ij}^{h^F}}{\partial {{r}_{ij}}} \mathbf{e}_{ij}$, 
where $\mathbf{r}_{ij} = \mathbf{r}_i - \mathbf{r}_j$, 
$h^F$ is smoothing length of fluid, 
and the unit vector $\mathbf{e}_{ij} = \frac{\mathbf{r}_{ij}}{r_{ij}}$, 
representing the derivative of the kernel function with respect to $\mathbf{r}_i$, 
the position of particle $i$.
Here, the inter-particle averaged velocity and pressure, 
$\widetilde{\mathbf{v}}{ij}$ and $\widetilde{p}{ij}$, 
can be evaluated using either a simple arithmetic average 
or a solution of the inter-particle Riemann problem. 
In this work, 
a low-dissipation Riemann solver is employed, 
and further details are provided in Ref. \cite{zhang2017weakly}.
The terms 
$\mathbf{f}_i^{S:p}\left(h^F\right)$ 
and $\mathbf{f}_i^{S:v}\left(h^F\right)$ 
denote the pressure and viscous forces, respectively, 
exerted on fluid particle $i$ by the solid structure.  

The SPH discretization of the solid conservation Eq. \eqref{momentum-conservation-s}, 
based on the initial configuration,
can be derived as
\begin{equation}\label{mom-sph}
	m_a \frac{\text{d} \mathbf{v}}{\text{d}t} = 2 \sum_b V_a V_b \tilde{\mathbb{P}}_{ab} \nabla_a^0 W_{ab}^{h^S} +\mathbf{g} + \mathbf{f}_a^{F:p}\left(h^F\right) +  \mathbf{f}_a^{F:v}\left(h^F\right),
\end{equation} 
where $\nabla_a^0 W_{a}^{h^S} = \frac{\partial W\left( \mathbf{r}_{ab}^0, h^S \right)}  {\partial r_{ab}^0} \mathbf{e}_{ab}^0$
with subscripts $a$ and $b$ denoting solid particles 
and $h^S$ the smoothing length of solid. 
Here, the inter-particle averaged first Piola-Kirchhoff stress $\tilde{\mathbb{P}}$ 
is defined as
\begin{equation}
	\tilde{\mathbb{P}}_{ab} = \frac{1}{2} \left( \mathbb{P}_a \mathbb{B}_a^0 + \mathbb{P}_b \mathbb{B}_b^0 \right), 
\end{equation}
where $\mathbb{B}^0_a = \left( \sum_b \left( \mathbf{r}_b^0 - \mathbf{r}_a^0 \right) \otimes \nabla_a^0 W_{ab} \right) ^{-1}$
is the correction matrix \cite{vignjevic2006sph, liu2010smoothed}.
Also $\mathbf{f}_a^{F:p}\left(h^F\right)$ and $\mathbf{f}_a^{F:v}\left(h^F\right)$ correspond to the fluid pressure 
and viscous forces acting on the solid particle $a$, respectively. 

In fluid-strucutre coupling, 
the surrounding solid structure is behaving as a moving boundary for fluid,
and the no-slip boundary condition is imposed at the fluid-structure interface.
Following Refs. \cite{adami2012generalized, zhang2021multi}, 
the interaction forces $\mathbf{f}_i^{S:p}\left(h^F\right)$ and $\mathbf{f}_i^{S:v}\left(h^F\right)$ acting on a fluid particle $i$, 
due to the presence of the neighboring solid particle $a$, 
can be obtained as
\begin{equation}\label{S:p}
	\mathbf{f}_i^{S:p}\left(h^F\right) = - 2  \sum_a V_i V_a \frac{p_i \rho^d_a+ p^d_a \rho_i}{\rho_i + \rho^d_a} \nabla_i W(\mathbf{r}_{ia}, h^F ), 
\end{equation}
and 
\begin{equation}\label{S:v}
	\mathbf{f}_i^{S:v}\left(h^F\right) = 2\sum_a  \eta V_i V_a \frac{\mathbf{v}_i - \mathbf{v}^d_a} {r_{ia}} \frac{\partial W(\mathbf{r}_{ia}, h^F )}{\partial {r_{ia}}} .
\end{equation}
Here, 
the imaginary pressure $p_a^d$ and velocity $\mathbf{v}_a^d$ 
of the solid particle
are defined by
\begin{equation}\label{fs-coupling}
	\begin{cases}
		p_a^d = p_i + \rho_i max(0, (\mathbf{g} - {\frac{\text{d}\mathbf{v}_a}{\text{d}t}}) \cdot \mathbf{n}^S) (\mathbf{r}_{ia} \cdot \mathbf{n}^S) \\
		\mathbf{v}_a^d = 2\mathbf{v}_i - \mathbf{v}_a
	\end{cases},
\end{equation}
where $\mathbf{n}^S$ denotes the surface normal direction of the solid structure. 
And the imaginary particle density $\rho_a^d$ is calculated through 
the EoS presented in Eq. \eqref{eq:eos}. 
Note that,
for the calculation of fluid-structure interaction forces, 
we use $h^F$ with assumption $h^F \geqslant h^S$. 
This will ensure that a fluid particle $i$ can be searched and tagged as a neighboring particle 
of a solid particle $a$ which is located in the neighborhood of particle $i$. 
Accordingly, the interaction forces $\mathbf{f}_a^{F:p}$ and $\mathbf{f}_a^{F:v}$ acting on a solid particle $a$ are given by
\begin{equation}\label{F:p}
	\mathbf{f}_a^{F:p}\left(h^F\right) = - 2  \sum_i V_a V_i \frac{p^d_a \rho_i +p_i \rho^d_a}{\rho_i + \rho^d_a} \nabla_a W(\mathbf{r}_{ai}, h^F ), 
\end{equation}
and 
\begin{equation}\label{F:v}
	\mathbf{f}_a^{F:v}\left(h^F\right) = 2\sum_i  \eta V_a V_i \frac{\mathbf{v}^d_a - \mathbf{v}_i} {r_{ai}} \frac{\partial W(\mathbf{r}_{ai}, h^F )}{\partial {{r}_{ai}}} .
\end{equation}
Note that the anti-symmetric property of the derivative of the kernel function will ensure the momentum conservation for each pair of interacting particles $i$ and $a$.  

\subsubsection{Time stepping}
The dual-criteria time-stepping method 
\cite{verlet1967computer, zhang2020dual, zhang2021multi} 
is employed to integrate the fluid equations. 
The fluid time step is constrained 
by an advection criterion $\Delta t_{ad}^F$ 
and an acoustic criterion $\Delta t_{ac}^F$, 
given by
\begin{equation}\label{dtf-advection}
	\Delta t_{ad}^F   =  0.25 \min\left(\frac{h^F}{|\mathbf{v}|_{max}}, \frac{{h^F}^2}{\eta}\right),
\end{equation}
and 
\begin{equation}\label{dt-relax}
	\Delta t_{ac}^F   = 0.6 \min \left( \frac{h^F}{c^F + |\mathbf{v}|_{max}} \right) .
\end{equation}
For the solid structure, the time-step criterion is
\begin{equation}\label{dts-advection}
	\Delta t^S   =  0.6 \min\left(\frac{h^S}{c^S + |\mathbf{v}|_{max}},
	\sqrt{\frac{h^S}{|\frac{\text{d}\mathbf{v}}{\text{d}t}|_{max}}} \right).
\end{equation}
In general, 
$\Delta t^S < \Delta t_{ac}^F$ because $c^S > c^F$. 
Rather than using $\Delta t^S$ as a single global time step for both phases, 
the structural dynamics are sub-cycled within each fluid acoustic step. 
Specifically, 
during one fluid acoustic time step $\Delta t_{ac}^F$, 
the structure is advanced $\kappa = \left\lfloor \frac{\Delta t_{ac}^F}{\Delta t^S} \right\rfloor + 1$ times, 
where $\lfloor \cdot \rfloor$ denotes the integer operation.

To eliminate the mismatch in force evaluation, 
the imaginary pressure $p_a^d$ and velocity $\mathbf{v}_a^d$ 
in Eq. \eqref{fs-coupling} are redefined as
\begin{equation} \label{fs-coupling-mr}
	\begin{cases}
		p_a^d = p_i + \rho_i max(0, (\mathbf{g} - \widetilde{\frac{\text{d} \mathbf{v}_a}{\text{d}t}}) \cdot \mathbf{n}^S) (\mathbf{r}_{ia} \cdot \mathbf{n}^S) \\
		\mathbf{v}_a^d = 2 \mathbf{v}_i  - \widetilde{\mathbf{v}}_a
	\end{cases}, 
\end{equation}
where, in Eq. \eqref{fs-coupling-mr}, 
$\widetilde{\mathbf{v}}_a$ and $\widetilde{\frac{d\mathbf{v}_a}{dt}}$
denote the time-averaged velocity and acceleration of the solid particles 
over one fluid acoustic time step.

A position-based Verlet scheme \cite{verlet1967computer, zhang2021multi} 
is employed. 
Quantities at the beginning, 
mid-point, and end of a fluid acoustic time step 
are denoted by the superscripts $n$, $n+\frac{1}{2}$, and $n+1$, respectively. 
The fluid integration is first performed as
\begin{equation}\label{verlet-first-half}
	\begin{cases}
		\rho_i^{n + \frac{1}{2}} = \rho_i^n + \frac{1}{2}\Delta t_{ac}^F  \frac{d \rho_i}{dt}\\
		\mathbf{r}_i^{n + \frac{1}{2}} = \mathbf{r}_i^n + \frac{1}{2} \Delta t_{ac}^F {\mathbf{v}_i}^{n}
	\end{cases}, 
\end{equation}
by advancing the density and position to the mid-point. 
The particle velocity is then updated to the new time level as
\begin{equation}\label{verlet-first-mediate}
	\mathbf{v}_i^{n + 1} = \mathbf{v}_i^{n} +  \Delta t_{ac}^F  \frac{d \mathbf{v}_i}{dt}. 
\end{equation}
Finally, the position and density of fluid particles are updated to the new time step by 
\begin{equation}\label{verlet-first-final}
	\begin{cases}
		\mathbf{r}_i^{n + 1} = \mathbf{r}_i^ {n + \frac{1}{2}} +  \frac{1}{2} \Delta t_{ac}^F {\mathbf{v}_i} \\
		\rho_i^{n + 1} = \rho_i^{n + \frac{1}{2}} + \frac{1}{2} \Delta t_{ac}^F \frac{d \rho_i}{dt}
	\end{cases}. 
\end{equation}
At this stage, 
the solid-particle acceleration induced by pressure and viscous forces 
is assumed to remain constant during the solid integration. 
For convenience, 
the index $\varkappa = 0, 1, \ldots, \kappa-1$ 
is introduced to denote the sub-steps of the solid time integration. 
Using the position-based Verlet scheme, 
the deformation tensor, density, and particle position 
are advanced to the mid-point as
\begin{equation}\label{verlet-first-half-solid}
	\begin{cases}
		\mathbb{F}_a^{\varkappa + \frac{1}{2}} = \mathbb{F}_a^{\varkappa} + \frac{1}{2} \Delta t^S \frac{\text{d} \mathbb{F}_a}{\text{d}t}\\
		\rho_a^{\varkappa + \frac{1}{2}} = \rho_a^0 \frac{1}{J} \\
		\mathbf{r}_a^{\varkappa + \frac{1}{2}} = \mathbf{r}_a^{\varkappa} + \frac{1}{2} \Delta t^S {\mathbf{v}_a}
	\end{cases}. 
\end{equation}
Then the velocity is updated by
\begin{equation}\label{verlet-first-mediate-solid}
	\mathbf{v}_a^{\varkappa + 1} = \mathbf{v}_a^{\varkappa} +  \Delta t^S  \frac{d \mathbf{v}_a}{dt}. 
\end{equation}
In the final stage, 
the deformation tensor and position of the solid particles 
are updated to the next solid time level as
\begin{equation}\label{verlet-first-final-solid}
	\begin{cases}
		\mathbb{F}_a^{\varkappa + 1} = \mathbb{F}_a^{\varkappa + \frac{1}{2}} + \frac{1}{2} \Delta t^S \frac{\text{d} \mathbb{F}_a}{\text{d}t}\\
		\rho_a^{\varkappa + 1} = \rho_a^0 \frac{1}{J} \\
		\mathbf{r}_a^{\varkappa + 1} = \mathbf{r}_a^{\varkappa + \frac{1}{2}} + \frac{1}{2} \Delta t^S {\mathbf{v}_a}^{\varkappa + 1}
	\end{cases}. 
\end{equation}
Before the next fluid step, 
the solid time integration described by Eqs.~\ref{verlet-first-half-solid}–\ref{verlet-first-final-solid} 
is performed for $\kappa$ sub-steps. 

Following Refs.~\cite{colagrossi2003numerical, zhang2017generalized}, 
the density is periodically reinitialized to prevent error accumulation 
from updating the density via the continuity equation 
in Eq.~\eqref{eq:mass_equation} during long-time simulations. 
Specifically, 
the free-surface density initialization scheme of Ref.~\cite{zhang2020dual}, 
accounting for solid particles within the support domain, 
is applied at the beginning of each fluid advection time step as follows:
\begin{equation}\label{density_initialization}
	\rho_i = \text{max} \left(\rho_i^*, \rho_i^0 
	\frac{\sum_j W(\mathbf{r}_{ij}, h^F ) 
		+ \sum_a W(\mathbf{r}_{ia}, h^F ) \frac{V_a^0}{V_i^0}}{\sigma^0}\right),
\end{equation}
where $\rho^*$ denotes the density before re-initialization, 
$\rho^0$ and $\sigma^0 = \sum_j W(\mathbf{r}_{ij}, h^F )$
are the initial density and the reference particle number density, respectively. 
For flows without free surface, 
Eq. \eqref{density_initialization} is simplified to 
\begin{equation}\label{density_initialization2}
		\rho_i = \rho_i^0 
		\frac{\sum_j W(\mathbf{r}_{ij}, h^F ) 
			+ \sum_a W(\mathbf{r}_{ia}, h^F ) \frac{V_a^0}{V_i^0}}{\sigma^0}
\end{equation}
which resets the density using the standard summation formulation \cite{monaghan1992smoothed}.
Note that $\frac{V_a^0}{V_i^0}$ is the 
initial volume ratio between the fluid and solid particle. 
With $\rho_i^0 = m_i \sigma^0$,
we have
\begin{equation}\label{density_initialization3}
		\rho_i 
		= m_i \sum_j W(\mathbf{r}_{ij}, h^F ) 
		+ \rho_i^0 \sum_a W(\mathbf{r}_{ia}, h^F ) V_a^0.
\end{equation}
To outline the implementation, 
the detailed simulation procedure is summarized in Algorithm~\ref{alg:a}.
{
	\linespread{1.0}\selectfont
	\begin{algorithm}[htb!]
		\caption{Multi-resolution SPH scheme for FSI problems.}
		\label{alg:a}
		\SetKwInOut{Input}{Input}
		\SetKwInOut{Output}{Output}
		Setup parameters and initialize the simulation\;
		Compute the initial number density of fluid particles\;
		Compute the surface normal direction of the structure\;
		Obtain the corrected configuration for the solid particles\;
		\While{the termination condition is not satisfied}
		{
			Compute the fluid advection time step $\Delta t_{ad}^F$\;
			Reinitialize the fluid density\;
			Update the surface normal direction of the structure\;
			\While{$\sum \Delta t_{ac}^F \leq \Delta t_{ad}^F$}
			{
				Compute the fluid acoustic time step $\Delta t_{ac}^F$\;
				Integrate the fluid equations using the position-based Verlet scheme\;
				Compute the pressure and viscous forces exerted on the structure by the fluid\;
				\While{$\sum \Delta t^S \leq \Delta t_{ac}^F$}
				{
					Compute the solid time step $\Delta t^S$\;
					Integrate the solid equations using the position-based Verlet scheme\;
				}
				Compute the time-averaged velocity and acceleration of the solid structure\;
			}
			Update the particle neighbor list, kernel values, and kernel gradients for the fluid particles\;
			Update the particle configuration between the fluid and solid particles\;
		}
		Terminate the simulation\;
	\end{algorithm}
}
%
\subsection{Fluid-shell interaction}
Building on the preliminary formulation, 
the fluid-induced pressure and viscous forces,
$\mathbf{f}_a^{F:p}\left(h^F\right)$ and $\mathbf{f}_a^{F:v}\left(h^F\right)$, 
also act on shell particle $a$ and contribute to its acceleration. 
Accordingly, the first conservation equation for the shell is discretized as
\begin{equation}\label{discrete_dynamic_equation1}
	\begin{split}
	m_{a}^0 \frac{d \mathbf{v}_{m, a}}{d t} =	&\sum\limits_b 
	\left(
	J_{m, a} \mathbb{N}_a
	\left(\mathbb{F}_{m, a}  \right)^{\operatorname{-T}} 
	\widetilde{\mathbb{B}}^{0, \bm{r}}_a 
	+ J_{m, b} \mathbb{N}_b \left(\mathbb{F}_{m, b}  \right)^{\operatorname{-T}}  
	\widetilde{\mathbb{B}}^{0, \bm{r}}_b
	\right)\\
	&\nabla^0 W_{ab} A_a^0 A_b^0 
	 +\mathbf{g} + \mathbf{f}_a^{F:p}\left(h^F\right) +  \mathbf{f}_a^{F:v}\left(h^F\right),
	\end{split}
\end{equation} 
where $\widetilde{\mathbb{B}}^{0, \bm{r}}$ is the correction matrix 
to remedy the 1st-order inconsistency \cite{wu2024sph}, 
and $A^0$ is the initial area of the shell particle. 
As the imaginary shell particles are introduced to retrieve the kernel completeness, 
the calculation of $\mathbf{f}_a^{F:p}\left(h^F\right)$ and $\mathbf{f}_a^{F:v}\left(h^F\right)$ on a shell particle $a$, 
and $\mathbf{f}_i^{S:p}\left(h^F\right)$ and $\mathbf{f}_i^{S:v}\left(h^F\right)$ on a fluid particle $i$, 
is needed to be modified accordingly as following sections. 
Note that only one-sided interaction is considered in this study.

\subsubsection{Generation of imaginary shell particles}
\begin{figure}[h]
	\centering
	\includegraphics[width=\textwidth]{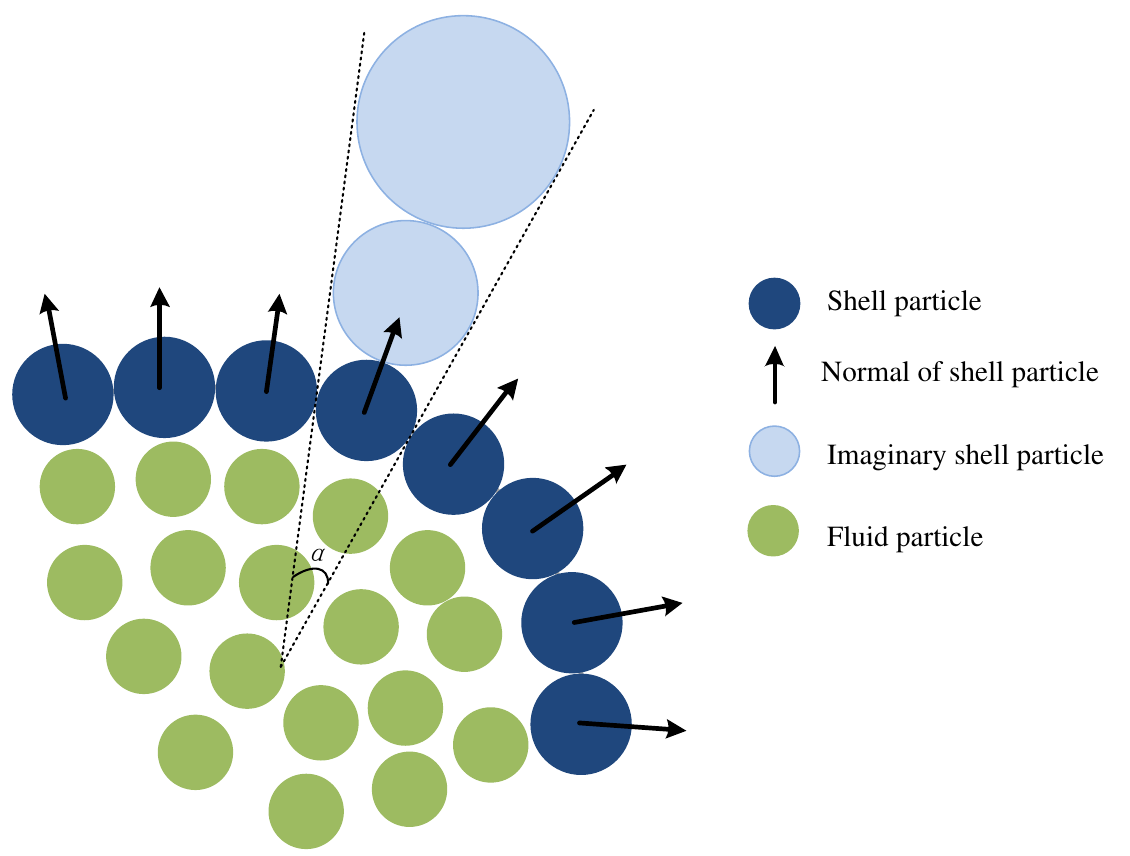}
	\caption{Setup of imaginary shell particles.}
	\label{figs:projection_shell}
\end{figure}

For a shell particle $a$ located in the support domain of a fluid particle $i$, 
i.e., $\left| \bm{r}_{ia} \right| \leq R^F$ with $R^F$ denoting the cut-off radius of fluid, 
the imaginary particles of this shell particle $a$ is considered, 
as shown in Figure \ref{figs:projection_shell}. 
The position of k-th imaginary particle is given as:
\begin{equation}
\bm{r}_a^k=\bm{r}_a + k \cdot dp^S \cdot \bm{n_a},
\end{equation} 
where $dp^S$ is the initial particle spacing of shell, 
$\bm{n}_a$ is the current normal direction of shell particle $b$, 
whose orientation is chosen to be consistent with the fluid-to-shell direction.

The area $A_a^k$ of an imaginary shell particle 
is determined from the local shell curvature.
As shown in Fig.~\ref{figs:projection_shell}. 
the curvature specifies the center of the local circle in 2D 
(or the local sphere in 3D). 
In 2D, two rays drawn from this center 
through the bounds associated with particle $a$ define the intercepted arc, 
whose length is taken as $A_a^k$. 
In 3D, the corresponding construction defines a spherical surface patch 
on the local sphere, 
and its area is taken as $A_a^k$. 
For the 2D case, $A_a^k$ is therefore given by
\begin{equation}\label{eq:area_calculation}
	A_a^k=\alpha (\frac{1}{\chi_a}+k \cdot dp^S), 
\end{equation} 
where $\alpha$ is the central angle of the circular sector shown in Figure \ref{figs:projection_shell}, 
which can be calculated as 
\begin{equation}\label{eq:area_calculation1}
	\alpha = A_a \cdot \chi_a.
\end{equation} 
Here, $\chi_a$ is the principle curvature of particle $a$, 
which can be calculated as 
\begin{equation}
	\chi_a =  \nabla \cdot \bm{n}_a.
\end{equation} 
Substituting Eq.\eqref{eq:area_calculation1} into Eq.\eqref{eq:area_calculation}, 
we have:
\begin{equation}\label{eq:area_calculation2}
	A_a^k=A_a (1+k \cdot \chi_a \cdot dp^S), 
\end{equation} 
For 3D problems, 
we can calculate the area with two principal curvatures, 
$\chi_{a1}$ and $\chi_{a2}$, 
and thus the area can be computed as:
\begin{equation}\label{eq:area_calculation3}
	A_a^k=A_a (1+k \cdot dp^S \cdot \chi_{a1}) (1+k\cdot dp^S \cdot \chi_{a2}). 
\end{equation} 
To get the curvatures $\chi_{a1}$ and $\chi_{a2}$, 
we first calculate the mean curvature $M$ and Gaussian curvature $N$ as
\begin{equation}
	M = \chi_{a1} + \chi_{a2} = \frac{1}{2} \nabla \cdot \bm{n}_a, 
\end{equation} 
\begin{equation}
	N = \chi_{a1} \cdot \chi_{a2} = 
	\frac{1}{2}\left[\left(\nabla \cdot \bm{n}_a\right)^2 - \sum_m \sum_n \left(\frac{\partial n_n}{\partial X_m}\frac{\partial n_m}{\partial X_n}\right)\right], \quad m, n  = {1, 2, 3}. 
\end{equation} 
Then $\chi_{a1}$ and $\chi_{a2}$ are written as
\begin{equation}
	\chi_{a1} = M + \sqrt{M^2 - N}, 
\end{equation} 
\begin{equation}
	\chi_{a2} = M - \sqrt{M^2 - N}.
\end{equation} 
Note that only if $\left| \bm{r}_{ia}^k \right| \leq R^F$
and the area $A_a^k$ is positive,
the k-th imaginary particle is applied.

\subsubsection{Equivalent value of kernel function}
\label{kernel_calculation}
Based on the preliminary formulation, 
the interaction forces are evaluated 
by Eqs.~\eqref{S:p} and \eqref{S:v} for the fluid phase 
and Eqs.~\eqref{F:p} and \eqref{F:v} for the solid phase, 
while the fluid density initialization is corrected 
in the presence of solids according to Eq.~\eqref{density_initialization}. 
The introduction of imaginary shell particles 
can be regarded as modifying the effective kernel quantities 
used in the above equations,
such that the underlying FSI coupling algorithm remains unchanged. 
The resulting equivalent kernel quantities are denoted by
$\overline W_{ia}$,  $\frac{\partial \overline W_{ia}}{\partial {{r}_{ia}}}$, 
$\frac{\partial \overline W_{ai}}{\partial {{r}_{ai}}}$, 
$\mathbf{\overline e}_{ia}$, and $\mathbf{\overline e}_{ai}$. 
Note that $\frac{\partial \overline W_{ia}}{\partial {{r}_{ia}}} 
= \frac{\partial \overline W_{ai}}{\partial {{r}_{ai}}}$, 
and $\mathbf{\overline e}_{ia} = -\mathbf{\overline e}_{ia}$.

In the density initialization Eq. \eqref{density_initialization2} of fluid 
with shell interaction, 
the contribution of a shell particle $a$ is calculated as:
\begin{equation}
	\sigma_a = \frac{\overline W(\mathbf{r}_{ia}, h^F ) V_a^0}{V_i^0} 
	= \frac{\overline W(\mathbf{r}_{ia}, h^F ) A_a^0 d_a^0}{V_i^0}
	= \frac{\sum_{k} W(\mathbf{r}_{ia}^k, h^F ) A_a^k dp^S}{V_i^0}, 
\end{equation} 
which means the equivalent kernel value 
$\overline W(\mathbf{r}_{ia}, h^F ) = \frac{1}{A_a^0 d_a^0} \sum \limits_k W(\mathbf{r}_{ia}^k, h^F ) A_a^k dp^S$, 
considering the effect of imaginary shell particles.

According to Eq. \eqref{S:v}, 
the viscous force $\mathbf{f}_i^{S:p}\left(h^F\right)$ 
acting on a fluid particle $i$ 
due to the presence of a shell particle $a$ 
is calculated as 
\begin{equation}\label{S_v_shell}
	\mathbf{f}_{ia}^{S:v}\left(h^F\right) = 2 \eta V_i \frac{\mathbf{v}_i - \mathbf{v}^d_a} {r_{ia}} \frac{\partial \overline W(\mathbf{r}_{ia}, h^F )}{\partial {r_{ia}}}V_a.
\end{equation}
While assuming that $\mathbf{v}^d_a$ and $r_{ia}$ 
are constant for each imaginary particle of shell particle $a$, 
Eq. \eqref{S_v_shell} can be written as
\begin{equation}\label{S_v_shell2}
	\mathbf{f}_{ia}^{S:v}\left(h^F\right) = 2 \eta V_i \frac{\mathbf{v}_i - \mathbf{v}^d_a} {r_{ia}} \sum_k \frac{\partial W(\mathbf{r}_{ia}^k, h^F )}{\partial {r_{ia}^k}} A_a^k dp^S.
\end{equation}
Comparing Eqs. \eqref{S_v_shell} and \eqref{S_v_shell2}, 
the equivalent kernel value $\frac{\partial \overline W(\mathbf{r}_{ia}, h^F )}{\partial {r_{ia}}}$ is 
\begin{equation}\label{S_v_shell3}
	\begin{split}
		\frac{\partial \overline W(\mathbf{r}_{ia}, h^F )}{\partial {r_{ia}}} 
		&= \frac{1}{V_a} \sum_k \frac{\partial W(\mathbf{r}_{ia}^k, h^F )}{\partial {r_{ia}^k}} A_a^k dp^S\\
		&= \frac{1}{A_a} \sum_k \frac{\partial W(\mathbf{r}_{ia}^k, h^F )}{\partial {r_{ia}^k}} A_a^k.
	\end{split}
\end{equation}

According to Eq. \eqref{S:p}, 
the pressure force $\mathbf{f}_i^{S:p}\left(h^F\right)$ 
acting on a fluid particle $i$ 
due to the presence of a shell particle $a$ 
is calculated as 
\begin{equation}\label{S_p_shell}
	\begin{split}
	\mathbf{f}_{ia}^{S:p}\left(h^F\right) 
	&= - 2 V_i \frac{p_i \rho^d_a+ p^d_a \rho_i}{\rho_i + \rho^d_a} \nabla_i \overline W(\mathbf{r}_{ia}, h^F ) V_a \\
	&= - 2 V_i \frac{p_i \rho^d_a+ p^d_a \rho_i}{\rho_i + \rho^d_a} 
	\frac{\partial \overline W(\mathbf{r}_{ia}, h^F )}{\partial {r_{ia}}}  \bm{\overline e}_{ia} V_a. 
\end{split}
\end{equation}
While assuming that $ p^d_a$
is constant for each imaginary particle of shell particle $a$, 
Eq. \eqref{S_v_shell} can be written as
\begin{equation}\label{S_p_shell2}
	\mathbf{f}_{ia}^{S:p}\left(h^F\right) = - 2 V_i \frac{p_i \rho^d_a+ p^d_a \rho_i}{\rho_i + \rho^d_a}
	\sum_k \frac{\partial W(\mathbf{r}_{ia}^k, h^F )}{\partial {r_{ia}^k}} 
	\bm{e}_{ia}^k A_a^k dp^S.
\end{equation}
Comparing Eqs. \eqref{S_p_shell} and \eqref{S_p_shell2},
and considering Eq. \eqref{S_v_shell3},
the equivalent value $\bm{\overline e}_{ia}$ is 
\begin{equation}\label{S_p_shell3}
		\bm{\overline e}_{ia}
		= \frac{\sum_k \frac{\partial W(\mathbf{r}_{ia}^k, h^F )}{\partial {r_{ia}^k}} \bm{e}_{ia}^k A_a^k}{\sum_k \frac{\partial W(\mathbf{r}_{ia}^k, h^F )}{\partial {r_{ia}^k}} A_a^k}.
\end{equation}
%
\subsection{Solid-shell interaction}
A particle-to-particle SPH contact model
is first formulated for solid–solid interactions by analogy to fluid dynamics, 
in which interpenetration is prevented.
Following the density initialization in Eq.~\eqref{density_initialization3}, 
a contact density is introduced for a solid particle $i$ 
in contact with another solid as
\begin{equation}\label{density_initialization4}
	\rho_i^c
	= \rho_i^0 \sum_a W(\mathbf{r}_{ia}, h^c ) V_a^0, 
\end{equation}
where the contact smoothing length is defined as
$h^c = \frac{h_i^c + h_a^c}{2}$, 
representing the arithmetic mean of the smoothing lengths 
of the two contacting solids, 
The contact particle $a$ is identified as a neighbor in the current configuration 
but not in the initial configuration, 
i.e., $r_{ia} \leq R^c$ and $r_{ia}^0 > R^c$. 

By analogy to the fluid momentum equation in Eq.~\eqref{fluid-model}, 
the contact interaction is then defined as
\begin{equation}\label{contact_model}
	m_i \frac{\text{d} \mathbf{v}^c_i}{\text{d} t}  
	=  - 2  \sum_a V_i V_a \widetilde{p}_{ia}^c \nabla_i W_{ia}^{h^c}
	= - 2  \sum_a V_i V_a \widetilde{p}_{ia}^c 
	\frac{\partial W\left( \mathbf{r}_{ia}, h^c \right)}  {\partial r_{ia}^0} \mathbf{e}_{ia},
\end{equation}
where $\widetilde{p}_{ia}^c = 0.5 \left(p_i^c + p_a^c\right)$ 
with $p_i^c = \rho_i^c \left(c_i^S\right)^2$
and $p_a^c = \rho_a^c \left(c_a^S\right)^2$ 
denoting the contact pressures.
Note that this formulation also takes into account the self-contact dynamics.

For shell-related contact, 
the Eqs. \eqref{density_initialization4} and \eqref{contact_model} 
are modified as 
\begin{equation}\label{density_initialization5}
	\rho_i^c
	= \rho_i^0 \sum_a \overline W(\mathbf{r}_{ia}, h^c ) V_a^0, 
\end{equation}
\begin{equation}\label{contact_model2}
	m_i \frac{\text{d} \mathbf{v}^c_i}{\text{d} t}  
	= - 2  \sum_a V_i V_a \widetilde{p}_{ia}^c 
	\frac{\partial \overline W\left( \mathbf{r}_{ia}, h^c \right)}  {\partial r_{ia}^0} \mathbf{\overline e}_{ia},
\end{equation}
where the equivalent values $\overline W(\mathbf{r}_{ia}, h^c )$, 
$\frac{\partial \overline W\left( \mathbf{r}_{ia}, h^c \right)}  {\partial r_{ia}^0}$
and $\mathbf{\overline e}_{ia}$ 
are same with those in section \ref{kernel_calculation}.

\section{Numerical examples}\label{sec:examples}
In this section, 
we conduct a series of benchmark tests with available analytical 
or numerical reference data from the literature 
to qualitatively and quantitatively assess the accuracy and stability of 
the proposed framework. 
Following the validation, 
we explore the deformation of a complex problem of oil tank car collision 
to showcase the potential of the present formulation. 
The $5th$-order Wendland kernel \cite{wendland1995piecewise}, 
characterized by a smoothing length of $h^S = 1.15 dp^S$ 
and a cut-off radius of $2.3 dp^S$, 
is employed for solid and shell dynamics throughout, 
while $h^F = 1.3dp^F$ for fluid dynamics.

\subsection{Hydrostatic FSI}\label{sec:hydrostatic}
In this section, 
the proposed FSI framework is validated using a hydrostatic benchmark test 
involving a water column resting on an elastic plate. 
The problem was originally proposed by Fourey et al.~\cite{fourey2017efficient} 
and has been widely used to assess 
the stability and accuracy of FSI solvers \cite{zhang2021multi, bao2024entirely}. 
In contrast to previous studies that model the plate as a full-dimensional solid 
\cite{khayyer2018enhanced, zhang2021multi, fourey2017efficient}, 
the plate here is represented as a thin shell 
using the formulation in Section~\ref{sec:interaction} 
and is coupled to the fluid through the proposed imaginary shell particle approach.

\begin{figure}[h]
	\centering
	\includegraphics[width=0.5\textwidth]{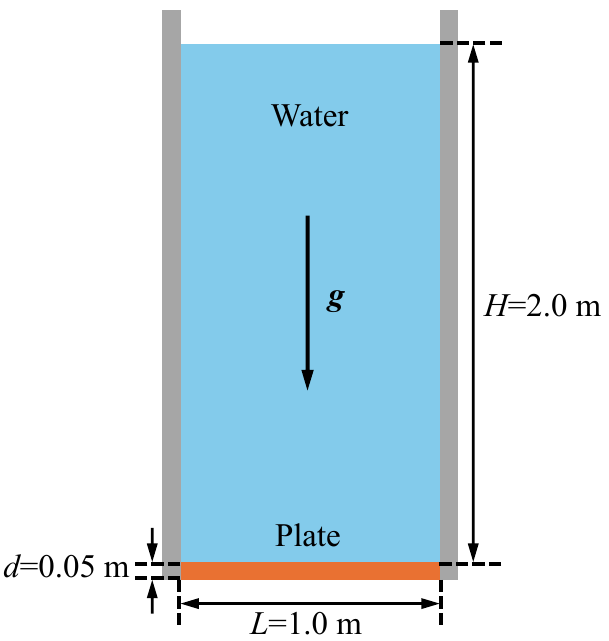}
	\caption{Hydrostatic FSI: Model setup.}
	\label{figs:hydro_FSI_setup}
\end{figure}

The problem setup is illustrated in Figure \ref{figs:hydro_FSI_setup}. 
A water column with a width of $L = 1.0 \, \text{m}$ 
and an initial height of $H = 2.0 \, \text{m}$ is contained in a tank. 
The bottom of the tank consists of a flexible aluminum plate, 
while the side walls are rigid. 
The fluid is modeled as water 
with a density of $\rho^F_0 = 1000.0 \, \text{kg/m}^3$ 
and a gravity of $g = 9.81 \, \text{m/s}^2$. 
The aluminum plate has a density of $\rho^S_0 = 2700.0 \, \text{kg/m}^3$,
a Young's modulus of $E = 67.5 \, \text{GPa}$, 
a Poisson's ratio of $\nu = 0.34$,
and a thickness of $d = 0.05 \, \text{m}$.
The plate is clamped at both ends. 
Initially, the system is at rest, 
and the plate deforms under the hydrostatic pressure of the water column.

The simulation is performed up to $t = 1.0,\text{s}$, 
by which time the initial oscillations have sufficiently damped 
and the system reaches a quasi-static equilibrium. 
According to the analytical solution in Ref.~\cite{fourey2017efficient}, 
the theoretical static mid-span deflection of the plate 
is $-6.85 \times 10^{-5},\text{m}$.

\begin{figure}[h]
	\centering
	\includegraphics[width=0.7\textwidth]{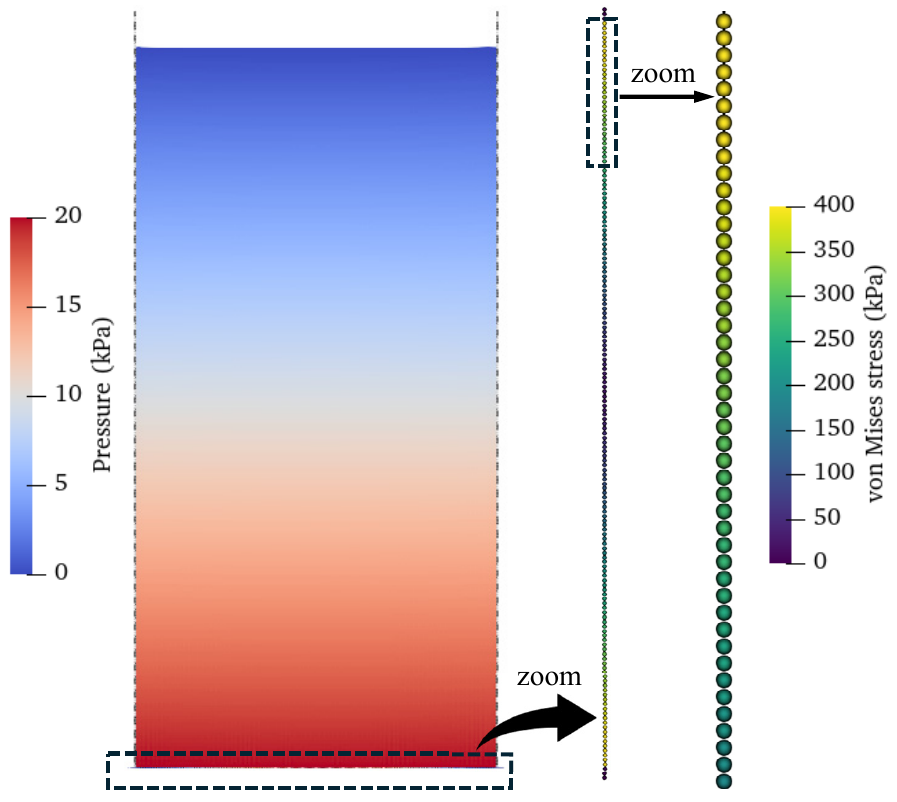}
	\caption{Hydrostatic FSI: The pressure distribution of the water column and the plate particles colored by von Mises stress.}
	\label{figs:hydro_FSI_pressure}
\end{figure}

Figure \ref{figs:hydro_FSI_pressure} shows the pressure field of the water column 
and the deformation of the shell plate at $t = 0.5 \, \text{s}$. 
It can be observed that the pressure field exhibits a smooth hydrostatic distribution 
without non-physical fluctuations near the fluid-shell interface. 
This demonstrates that the proposed imaginary shell particle method 
effectively completes the kernel support for fluid particles near the boundary, 
preventing particle penetration and ensuring correct pressure transmission.
Simultaneously,
the von Mises stress distribution on the shell particles 
also exhibits a smooth pattern without introducing any numerical instabilities.

\begin{figure}[h]
	\centering
	\includegraphics[width=0.6\textwidth]{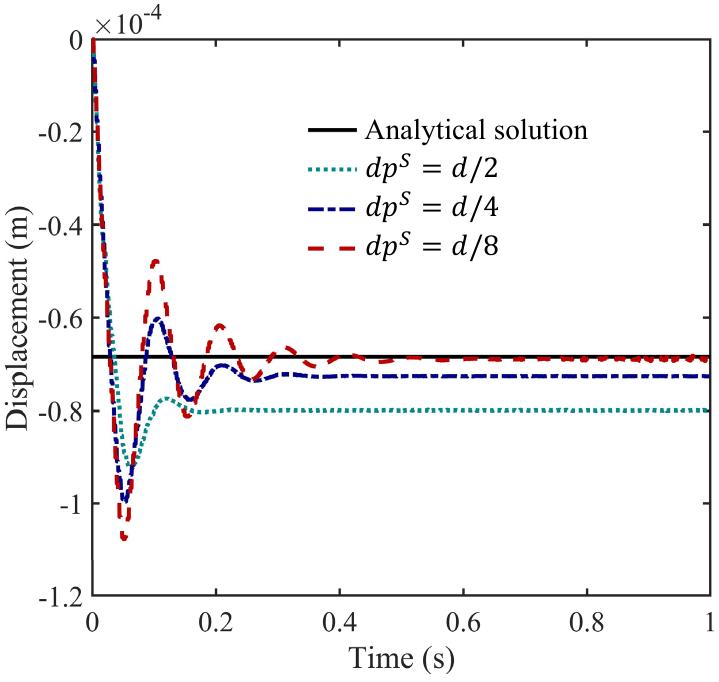}
	\caption{Hydrostatic FSI: Time history of the vertical
	mid-span displacement of the elastic plate.}
	\label{figs:hydro_FSI_disp}
\end{figure}

The time history of the vertical displacement at the mid-span of the plate 
is plotted in Figure \ref{figs:hydro_FSI_disp}. 
The result shows that the plate oscillates initially and gradually converges to a stable value. 
Three different particle resolutions ($dp = dp^F = dp^S = d/2$, $d/4$ and $d/8$) 
are examined to assess the convergence behavior of the method. 
With the increase of resolution, 
the oscillation amplitude decreases, 
and the converged displacement approaches the analytical solution.
The converged displacement obtained by the present shell-based SPH method 
agrees very well with the analytical solution ($-6.85 \times 10^{-5} \, \text{m}$) 
and the numerical results from solid-based SPH solvers \cite{zhang2021multi}.
This validates that the proposed framework 
can accurately capture the hydro-elastic response, 
proving that the reduced-dimensional shell model 
along with the projection method, 
effectively mimics the behavior of a full-dimensional solid boundary
in hydrostatic FSI scenarios.

\subsection{Dam-break flow through an elastic gate}\label{sec:elastic_gate}
To further evaluate the performance of the proposed framework 
in handling free-surface flows interacting with thin flexible structures, 
the benchmark problem of a dam-break flow through an elastic gate is simulated. 
This problem was originally investigated experimentally 
by Antoci et al. \cite{antoci2007numerical} 
and has been widely adopted as a standard validation case for various FSI solvers 
\cite{khayyer2018enhanced, zhang2021multi, fourey2017efficient}.
It is important to note that in these reference studies, 
the elastic gate was typically modeled using solid-based discretization methods, 
such as multi-layer SPH solid particles \cite{khayyer2018enhanced, zhang2021multi} 
or solid finite elements \cite{fourey2017efficient}. 
In contrast, 
the present work utilizes the proposed shell formulation
represented by a single layer of particles to model the gate, 
which significantly reduces the computational cost because of less particles
while maintaining accuracy.

\begin{figure}[htb!]
	\centering
	\includegraphics[width=0.6\textwidth]{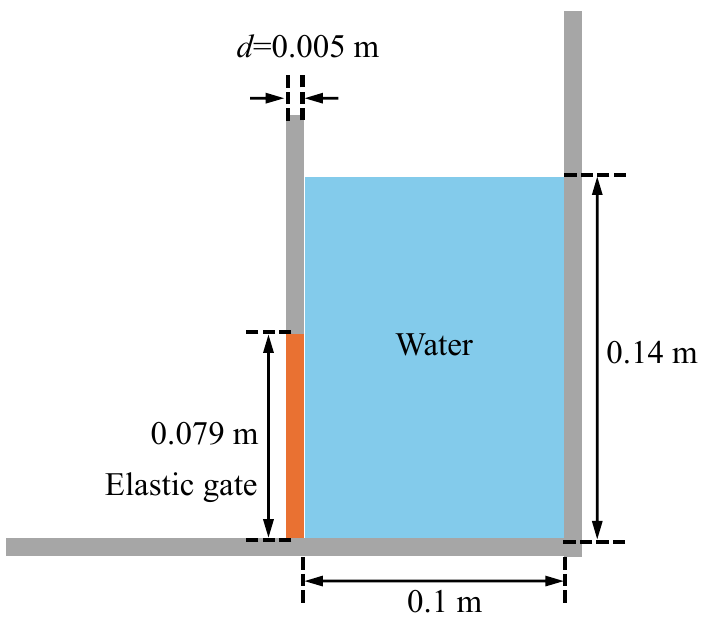}
	\caption{Dam-break flow through an elastic gate: Initial configuration and geometric parameters.}
	\label{figs:elastic_gate_setup}
\end{figure}

The computational setup is illustrated in Figure \ref{figs:elastic_gate_setup}. 
A tank contains a water column with an initial height of $0.14 \, \text{m}$ 
and a width of $0.1 \, \text{m}$. 
An elastic gate, 
acting as a dam, 
is clamped at its upper end to a rigid wall, 
while its lower end is free to move. 
The gate has a height of $0.079 \, \text{m}$ and a thickness of $d = 0.005 \, \text{m}$. 
The fluid is modeled as water with a density of $\rho^F_0 = 1000.0 \, \text{kg/m}^3$. 
The elastic gate is made of rubber with a density of $\rho^S_0 = 1100.0 \, \text{kg/m}^3$, 
a Young's modulus of $E = 7.8 \, \text{MPa}$, 
and a Poisson's ratio of $\nu = 0.47$.

\begin{figure}[htb!]
	\centering
	\includegraphics[width=\textwidth]{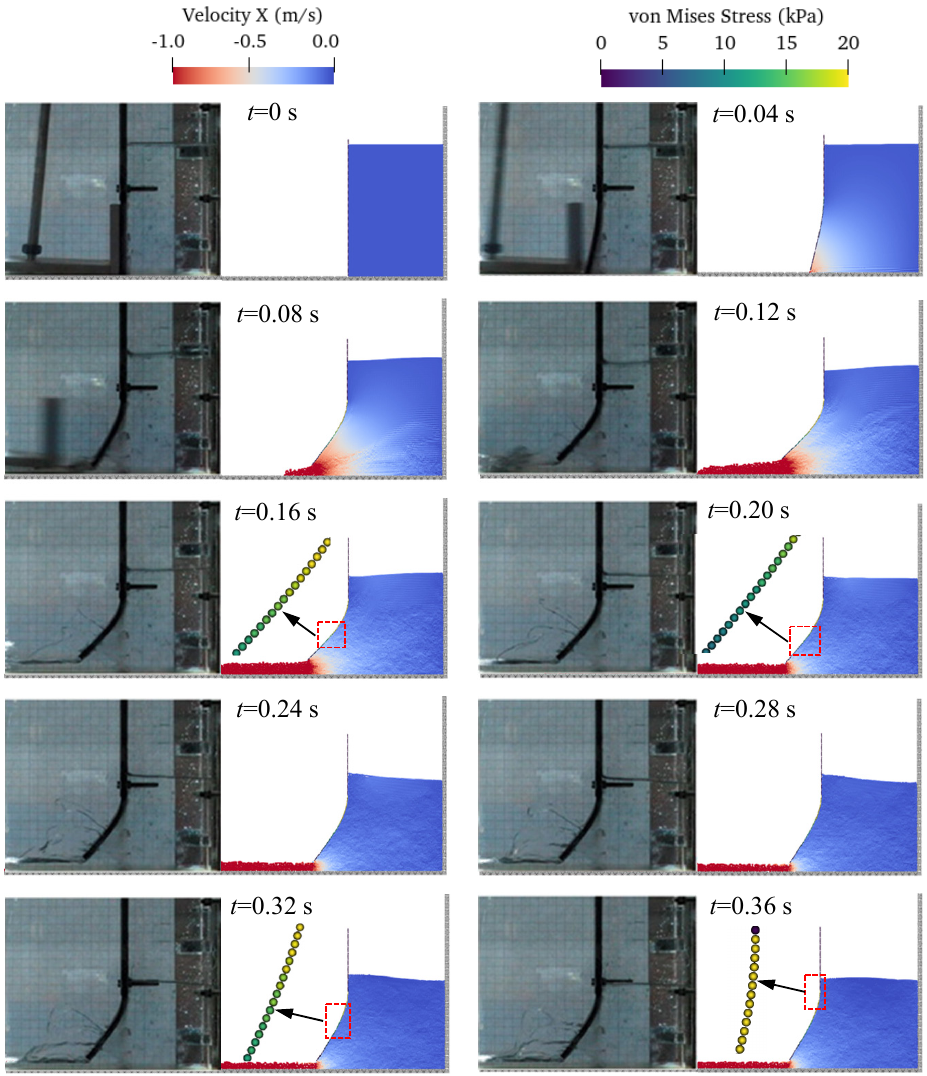}
	\caption{Dam-break flow through an elastic gate: Snapshots of the fluid velocity field and the von Mises stress distribution on the shell structure at different time instants. The deformed shapes are compared with the experimental profiles from Antoci et al. \cite{antoci2007numerical}.}
	\label{figs:elastic_gate_pressure}
\end{figure}

At $t=0 \, \text{s}$, the water column is released under gravity ($g = 9.81 \, \text{m/s}^2$). 
The hydrostatic pressure acts on the elastic gate, 
inducing significant non-linear deformation 
and allowing the water to flow underneath.
Figure \ref{figs:elastic_gate_pressure} presents snapshots of the simulation 
at representative time instants. 
The fluid domain is colored by horizontal velocity, 
while the shell particles are colored by the von Mises stress. 
Qualitatively, 
the predicted deformation profiles of the gate 
and the flow patterns show strong agreement with the experimental observations 
by Antoci et al. \cite{antoci2007numerical}.

\begin{figure}[htb!]
	\centering
	\includegraphics[width=1.0\textwidth]{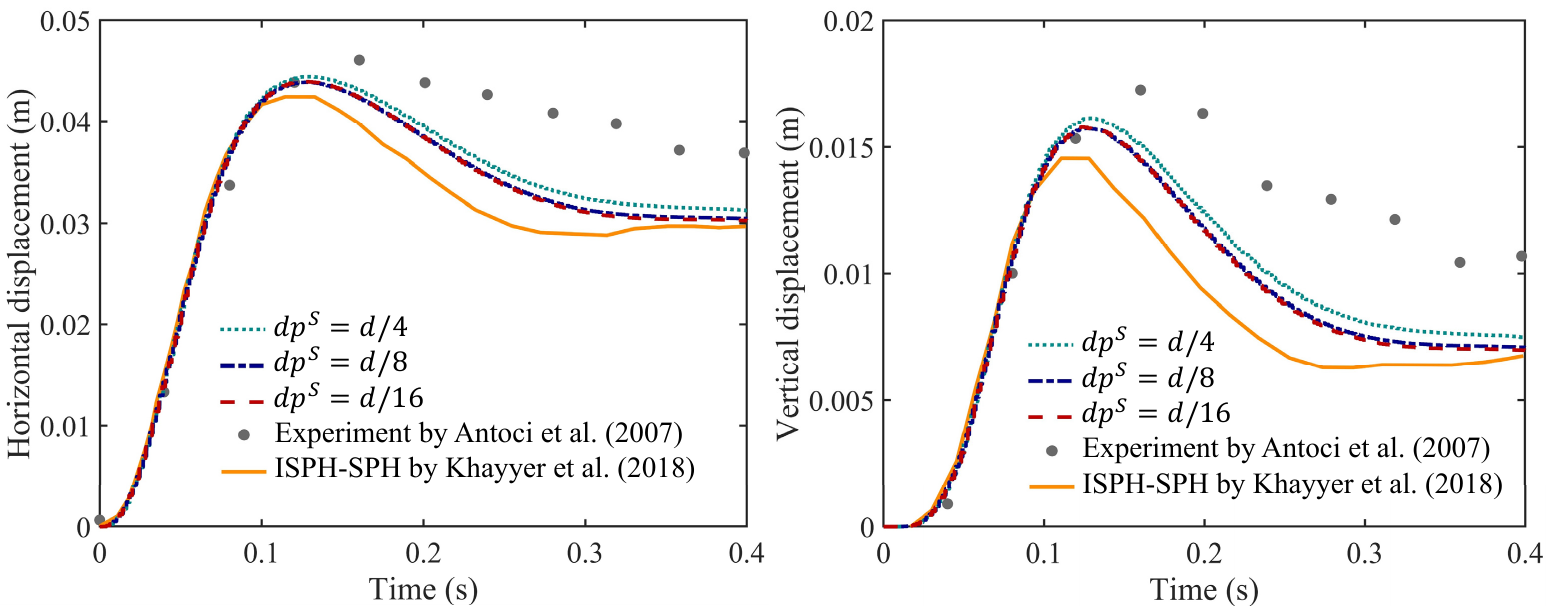}
	\caption{Dam-break flow through an elastic gate: Time histories of the horizontal (left) and vertical (right) displacements of the free end of the gate. The present results with three resolutions ($dp=d/4, d/8, d/16$) are compared with the experimental data of Antoci et al. \cite{antoci2007numerical} and the numerical results of Khayyer et al. \cite{khayyer2018enhanced}.}
	\label{figs:elastic_gate_disp}
\end{figure}

To quantitatively validate the accuracy and convergence of the shell-based approach, 
the horizontal and vertical displacements of the gate's free end are monitored. 
A convergence study is conducted using three different particle resolutions: 
$dp = dp^F = dp^S = d/4$, $d/8$, and $d/16$, 
where $d$ is the thickness of the elastic gate. 
The time histories of the tip displacements 
are plotted in Figure \ref{figs:elastic_gate_disp}. 
A clear convergence trend is evident:
as the particle resolution increases, 
the difference between the results of adjacent resolutions 
(i.e., between $d/4$ and $d/8$, and between $d/8$ and $d/16$) 
progressively diminishes, 
indicating that the solution is approaching a resolution-independent state.

Furthermore, 
the trade of the displacement curves exhibit good agreement 
with both the experimental data \cite{antoci2007numerical} 
and the numerical results
 from the enhanced ISPH-SPH method \cite{khayyer2018enhanced}.
It is observed that some deviations exist between the simulation 
and the experiment during the closing phase of the gate (the rebound period).
Consistent with findings in literature 
\cite{zhang2021multi, zhang2019smoothed, khayyer2018enhanced,yang2012free}, 
these discrepancies are likely attributed to the simplified material model 
employed in the simulation \cite{yang2012free}, 
which may not fully capture the complex hysteretic damping properties 
of the actual rubber material used in the experiment.
Despite this, 
the proposed fluid-shell coupling framework 
proves capable of reproducing the dominant physics of FSI scenarios 
with accuracy comparable to fluid-solid models.

\subsection{Dam-break impact on an elastic plate}\label{sec:Dambreak}
To rigorously validate the proposed FSI framework for hydroelastic problems 
involving violent free-surface flows and large structural deformations, 
the benchmark case of a 3D dam-break flow impacting an elastic plate 
is simulated. 
The numerical setup faithfully reproduces the experimental configuration 
conducted by Liao et al. \cite{liao2015free},
and the results are compared with both experimental data and numerical results 
from previous studies \cite{sun2019study, sun2021accurate, zhang2023efficient}.

\begin{figure}[h]
	\centering
	\includegraphics[width=0.7\textwidth]{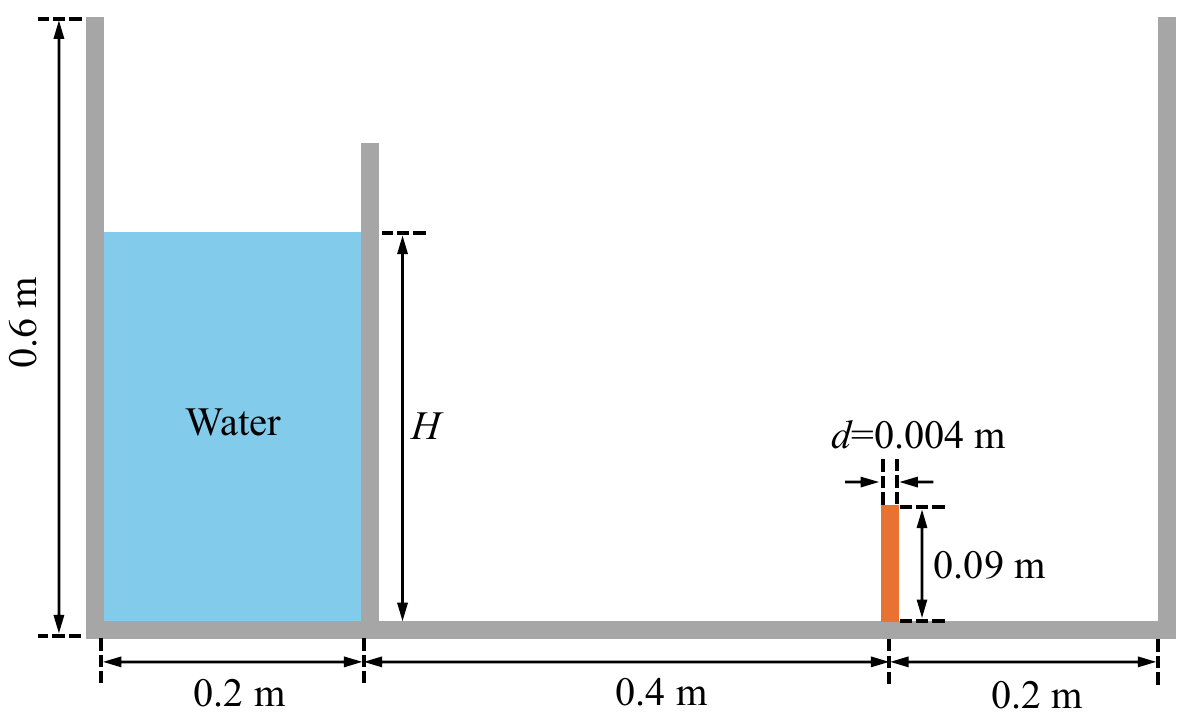}
	\caption{Dam-break flow impacts an elastic plate: 2D schematic diagram of the FSI dam-breaking test setup. The 3D geometry has an identical cross-section with a width of $0.2 \, \text{m}$.}
	\label{figs:dambreak_plate_setup}
\end{figure}

The geometric setup is shown in Fig.~\ref{figs:dambreak_plate_setup}. 
A water column is initially confined in a tank by a vertical gate. 
Downstream of the water column, 
a vertical elastic plate is installed and clamped at the tank bottom. 
In this study, 
the case with an initial water height of $H = 0.2,\text{m}$ 
is considered for validation. 
The plate is made of rubber with density $\rho_0 = 1164.0,\text{kg/m}^3$, 
Young’s modulus $E = 3.5,\text{MPa}$, 
and Poisson’s ratio $\nu = 0.49$, 
and has a thickness of $d = 0.004,\text{m}$. 
To quantify the structural response, 
the horizontal displacement is recorded at a marker on the plate centerline 
located $0.0875,\text{m}$ above the bottom, 
consistent with the experimental measurement point in Ref.~\cite{liao2015free}.
\begin{figure}[htb!]
	\centering
	\includegraphics[width=\textwidth] {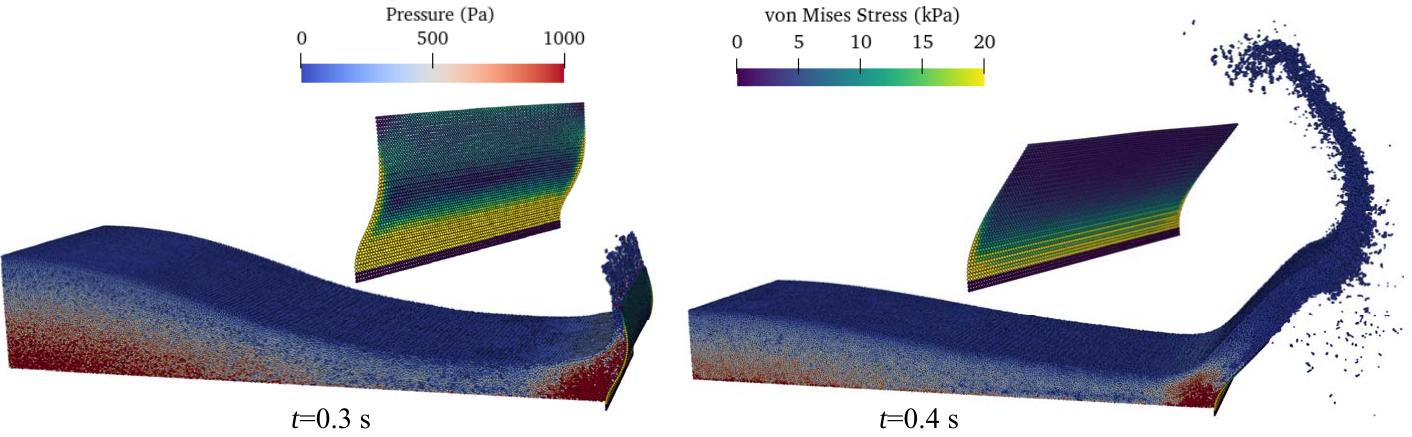}
	\caption{Dam-break flow impacts an elastic plate: 3D SPH simulation snapshots showing the fluid distribution and structural deformation at $t=0.3 \, \text{s}$ and $t=0.4 \, \text{s}$.}
	\label{figs:dambreak_3D_snapshots}
\end{figure}
\begin{figure}[htb!]
	\centering
	\includegraphics[width=0.7\textwidth] {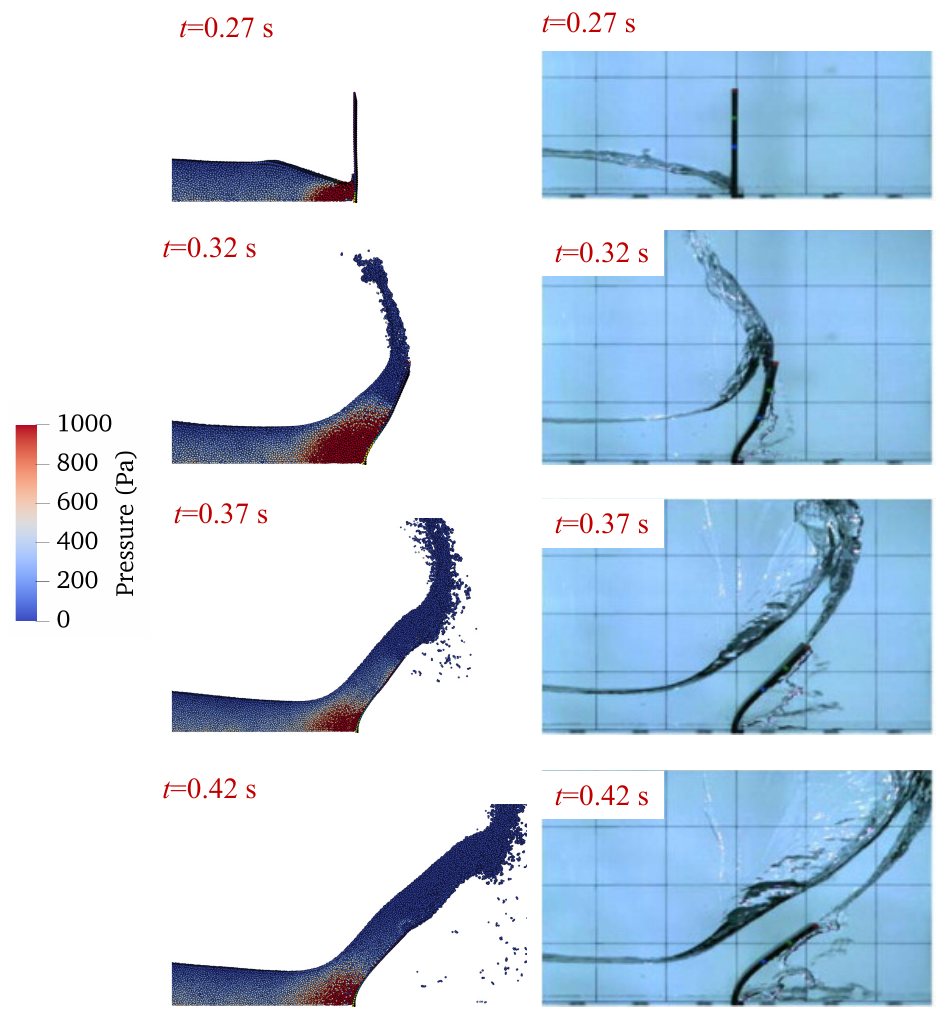}
	\caption{
		Dam-break flow impacts an elastic
		plate: Comparison between the present SPH results and experimental results \cite{liao2015free} at four typical time instants.}
	\label{figs:dambreak_experiment}
\end{figure}
\begin{figure}[htb!]
	\centering
	\includegraphics[width=0.7\textwidth] {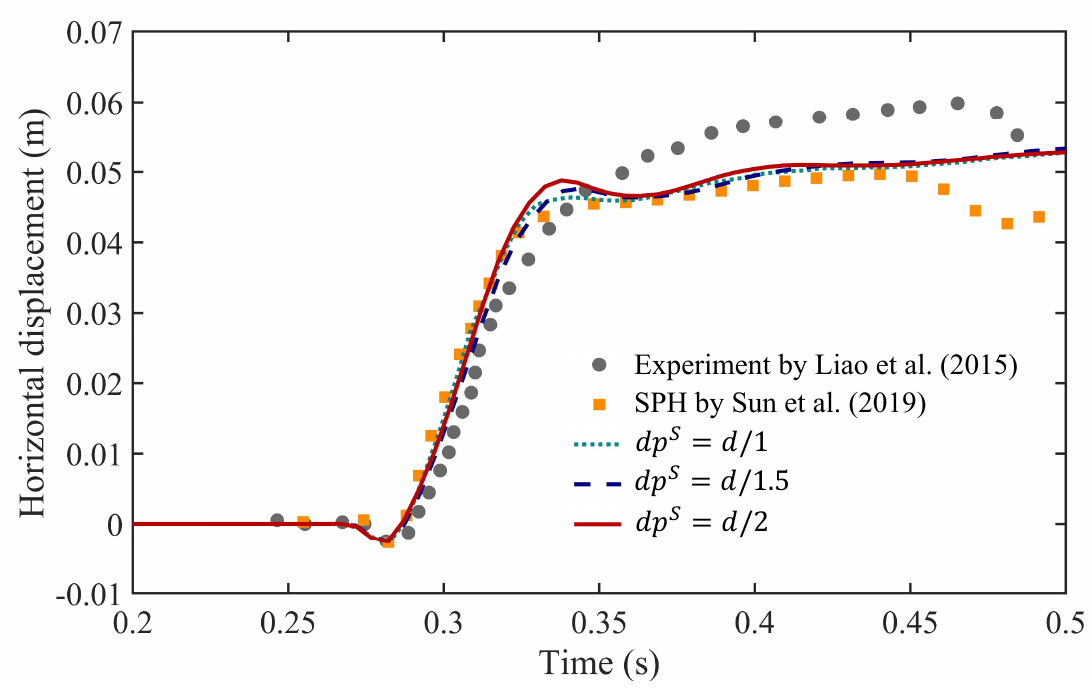}
	\caption{
		Dam-break flow impacts an elastic plate: 
		Time history of the horizontal displacements
		on the elastic plate for the case 
		with an initial water depth of $H = 0.2 , \text{m}$ 
		under three spatial particle discretization. 
		The SPH results are compared with the experimental data from Ref. \cite{liao2015free} and SPH result from Ref. \cite{sun2019study}.}
	\label{figs:dambreak_disp}
\end{figure}

As highlighted by Sun et al. \cite{sun2019study}, 
the kinematics of the gate removal plays a critical role 
in the early development of the flow and the subsequent impact forces. 
To minimize discrepancies arising from the instant removal assumption, 
the gate motion in the SPH simulation is prescribed 
using the following polynomial law 
derived from experimental observations \cite{sun2019study}:
\begin{equation}
    h_g(t) = -285.115 t^3 + 72.305 t^2 + 0.1463 t,
\end{equation}
where $h_g(t)$ denotes the vertical position of the gate. 
The simulation is conducted over a physical duration of $t = 0.5,\text{s}$. 
As reported in Ref.~\cite{sun2019study}, 
single-phase simulations typically start to deviate 
from the experimental measurements beyond this time 
because an air cavity becomes entrapped behind the plate. 
The resulting air-cushion effect markedly affects the rebound dynamics 
and would require a multiphase air–water–structure model, 
which is beyond the scope of the present study.

Figure \ref{figs:dambreak_3D_snapshots} illustrates 
the 3D spatiotemporal evolution of the water-plate interaction. 
The proposed framework successfully captures the complex flow features, 
including the run-up, splashing, 
and the interplay between fluid loading and structural deformation. 
Specifically, 
at $t=0.3 \, \text{s}$, 
the impinging flow surges upward upon contact with the barrier. 
The plate responds with a synchronized deformation, 
adopting a characteristic "S-shaped" profile. 
This deformation mode occurs 
because the hydrodynamic impact momentum 
is primarily concentrated on the lower-central region of the structure at this stage. 
Subsequently, 
by $t=0.4 \, \text{s}$, 
the fluid is guided upward along the deflecting structure, 
generating significant splashing. 
As the hydrodynamic loading redistributes to the upper portion of the plate, 
the deformation mode transitions into a uniform arc shape.

A qualitative comparison is presented in Figure \ref{figs:dambreak_experiment}, 
where the numerical snapshots 
are aligned with experimental photographs \cite{liao2015free} 
at four typical time instants. 
The SPH results show good agreement with the experiment 
in terms of the free surface profile and the structural deflection. 
To further assess the accuracy and convergence of the method, 
a spatial resolution study is conducted using three different particle spacings: 
$dp = dp^F = dp^S = d$, $d/1.5$, and $d/2$, where $d$ is the plate thickness. 
The time histories of the horizontal displacement at the marker point 
are plotted in Figure \ref{figs:dambreak_disp}. 
The results align well with both the experimental data \cite{liao2015free} 
and the reference numerical solution \cite{sun2019study}. 
This confirms the capability of the proposed multi-resolution framework 
to accurately predict the fluid-structure interaction dynamics in 3D space.
\subsection{Flow around cylinder}\label{sec:flow_around_cylinder}
\begin{figure}[h]
	\centering
	\includegraphics[width=0.8\textwidth]{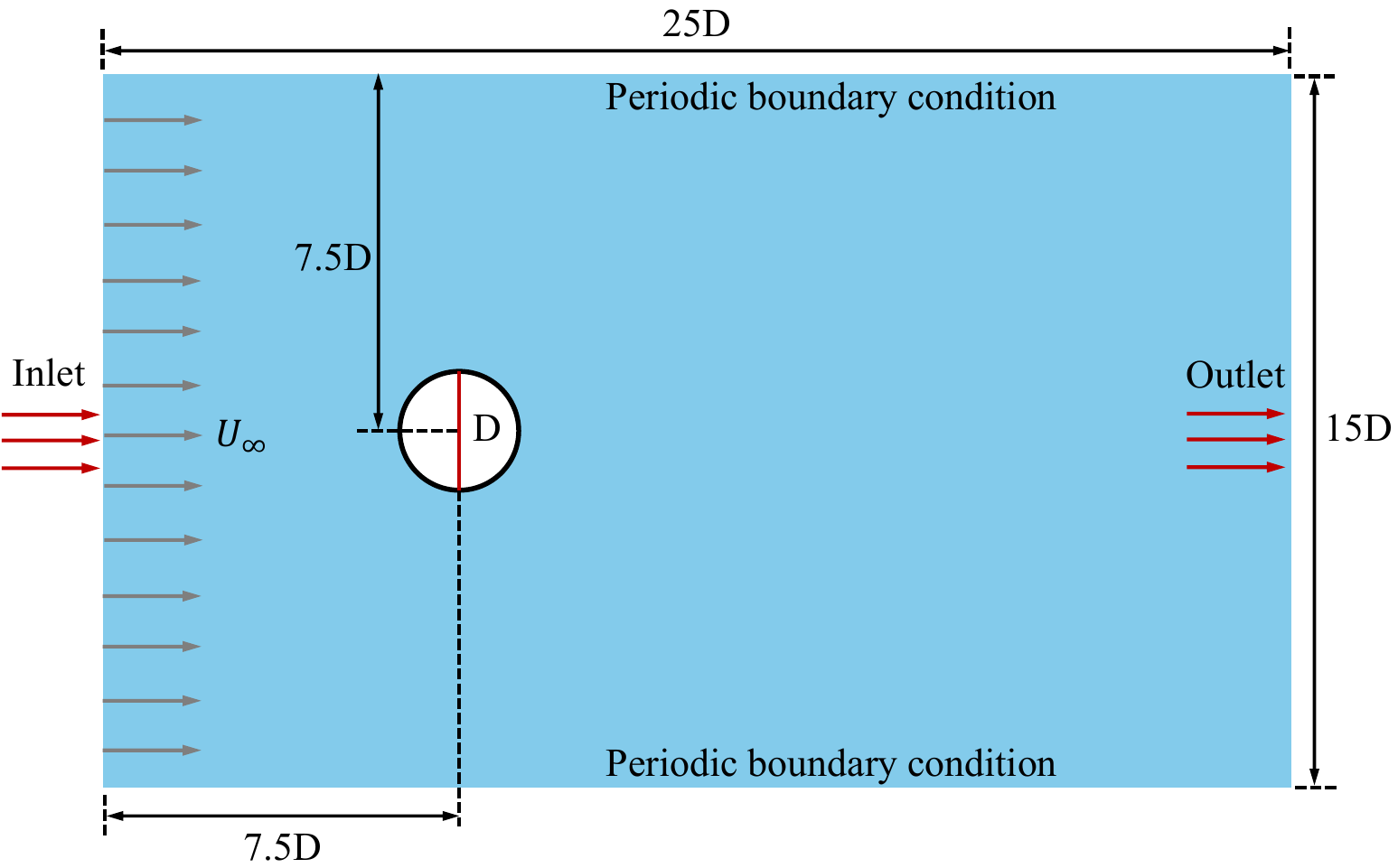}
	\caption{Flow around cylinder: Model setup.}
	\label{figs:cylinder_setup}
\end{figure}

To evaluate the capability in capturing unsteady wake dynamics, 
we consider the canonical benchmark of two-dimensional laminar flow 
past a circular cylinder.
In the present study, 
the cylinder is modeled as a rigid shell structure 
interacting with the surrounding fluid. 
The simulation is conducted 
at a Reynolds number of $Re = \rho_{\infty} U_{\infty} D / \mu = 100$. 
The fluid density and velocity in the far-field are set to $\rho_{\infty} = 1.0$ 
and $U_{\infty} = 1.0$, 
and the cylinder diameter $D = 2.0$. 
To minimize boundary effects, 
a rectangular computational domain of size $[25D, 15D]$ is employed, 
with the cylinder center located at $(7.5D, 7.5D)$. 
Periodic boundary conditions are applied to both the streamwise (left-right) 
and transverse (top-bottom) boundaries. 
The simulation is integrated up to a final physical time of $t = 300$ 
to ensure a fully developed periodic flow state. 
The model setup, 
highlighting the shell-based representation of the cylinder, 
is illustrated in Figure \ref{figs:cylinder_setup}.

\begin{figure}[h]
	\centering
	\includegraphics[width=1\textwidth]{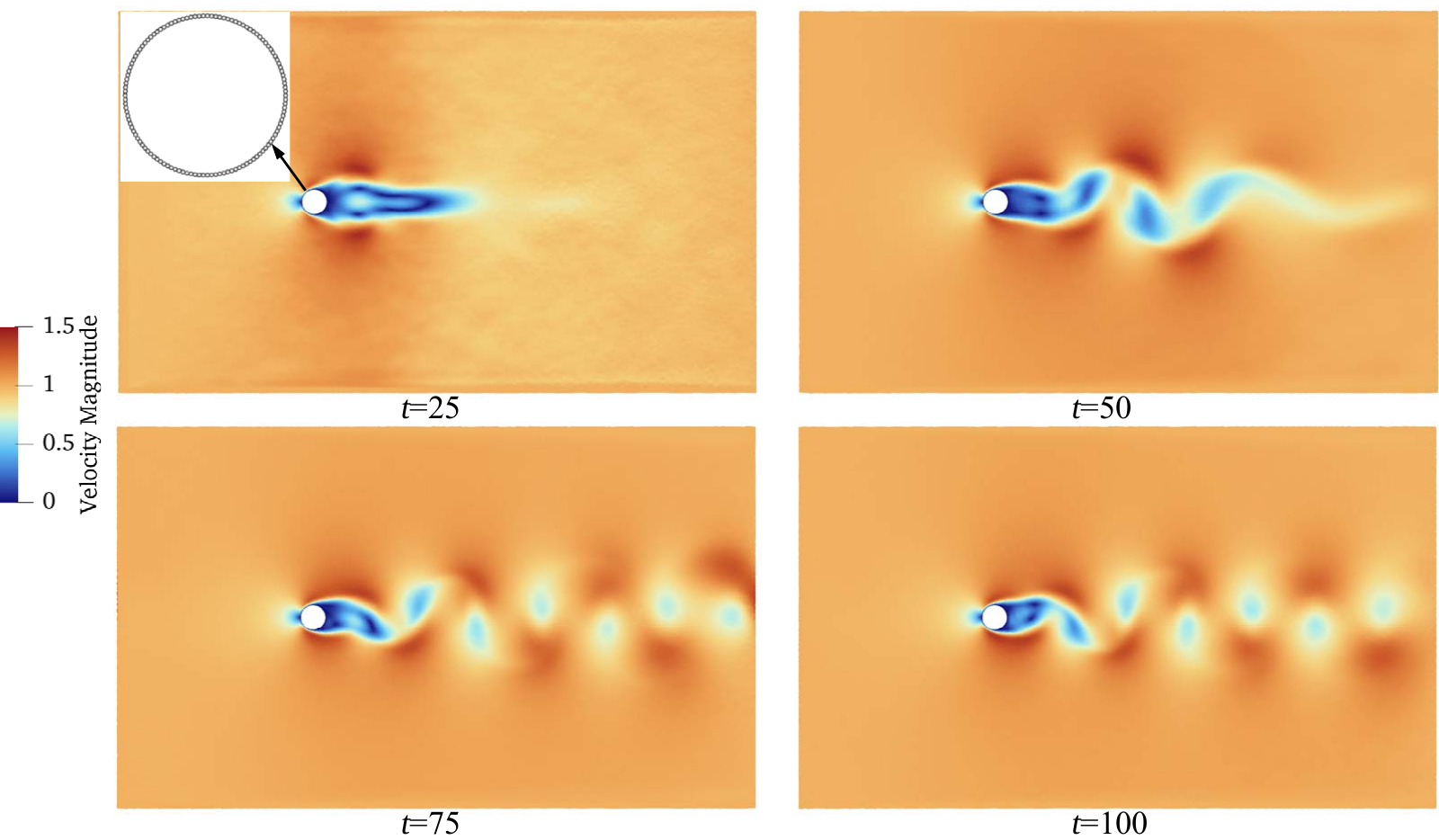}
	\caption{Flow around cylinder: Velocity contour at different time instants. $Re=$ 100 and $dp=$ 0.05.}
	\label{figs:cylinder_velocity}
\end{figure}

\begin{figure}[h]
	\centering
	\includegraphics[width=1\textwidth]{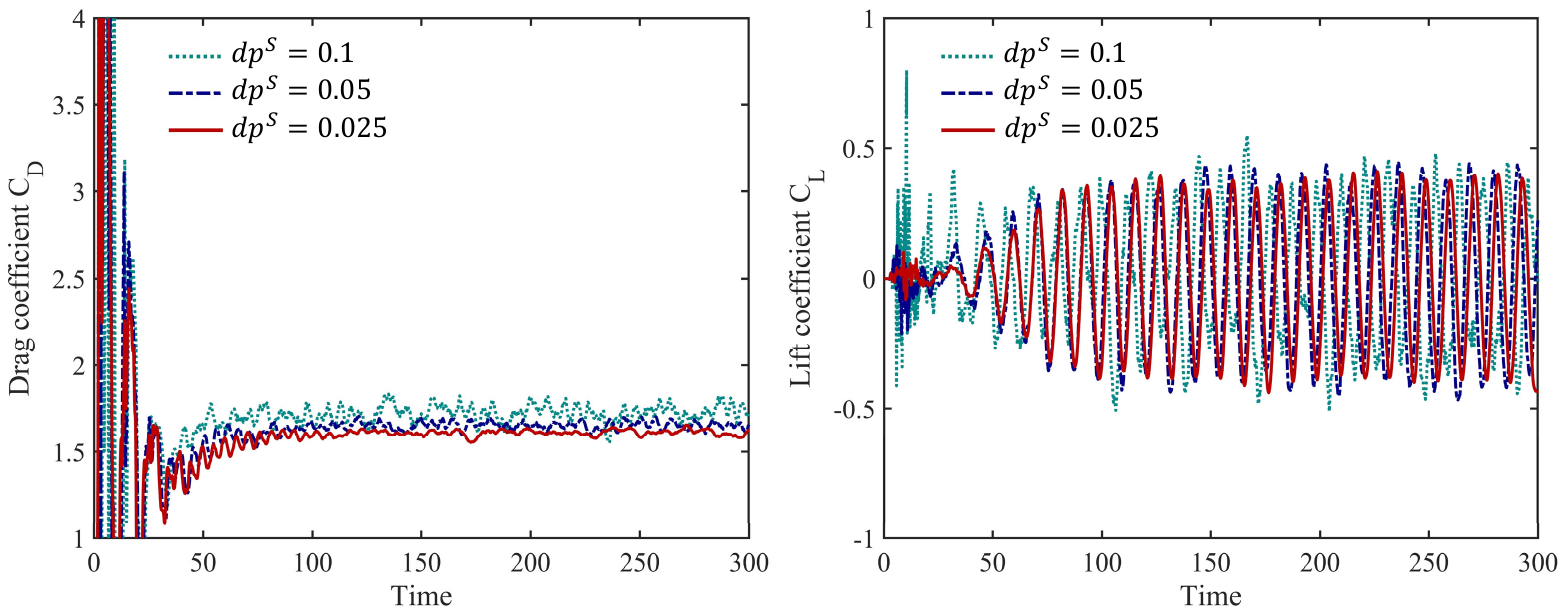}
	\caption{Flow around a cylinder: Drag (left panel) and lift (right panel) coefficients with the time for different particle reolutions ($Re=$ 100).}
	\label{figs:cylinder_Cd}
\end{figure}

\begin{table}[htbp]
    \centering
	\small
    \caption{Flow around a cylinder: Comparison of drag and lift coefficients and Strouhal number with reference data at $Re=100$ and $dp=0.025$.}
    \label{tab:validation}
    \begin{tabular}{lcccc}
        \toprule
         &  $C_D$ & $C_L$ & $St$ \\
        \midrule
        White \cite{white2006viscous} & 1.46 & -- & -- \\
        Brehm et al. \cite{brehm2015locally} & 1.32$\pm 0.0100$ & $\pm 0.320$ & 0.165 \\
        Al-Marouf and Samtaney \cite{al2017versatile} & 1.34$\pm 0.0089$ & $\pm 0.325$ & 0.166 \\
        Liu et al. \cite{liu1998preconditioned} & 1.35$\pm 0.0120$ & $\pm 0.339$ & 0.165 \\
        Le et al. \cite{le2006immersed} & 1.37$\pm 0.0090$ & $\pm 0.323$ & 0.160 \\
        Russell and Wang \cite{russell2003cartesian}  & 1.38$\pm 0.0070$  & $\pm 0.300$ & 0.172 \\
        Zhang et al. \cite{zhang2023lagrangian}  & 1.61$\pm 0.0050$ & $\pm 0.448$ & 0.171 \\
		Present  & 1.59$\pm 0.0226$ & $\pm 0.421$ & 0.179 \\
        \bottomrule
    \end{tabular}
\end{table}

The hydrodynamic forces acting on the cylinder 
are quantified using the drag ($C_D$) and lift ($C_L$) coefficients, 
defined as:
\begin{equation}
    C_D = \frac{F_D}{\frac{1}{2}\rho_{\infty} U_{\infty}^2 D}, \quad C_L = \frac{F_L}{\frac{1}{2}\rho_{\infty} U_{\infty}^2 D},
\end{equation}
where $F_D$ and $F_L$ represent the total fluid forces 
in the streamwise and transverse directions, respectively. 
The periodicity of the wake is characterized by the Strouhal number, 
$St = f D U_{\infty}^{-1}$, 
where $f$ denotes the vortex shedding frequency 
derived from the spectral analysis of the lift coefficient.

Figure \ref{figs:cylinder_velocity} visualizes the velocity magnitude field 
at representative instants. 
The results distinctively capture the shear layer separation 
from the cylinder surface 
and the subsequent formation of the von Kármán vortex street, 
characterized by the alternating shedding of counter-rotating vortices downstream.

The temporal evolution of the coefficients 
is presented in Figure \ref{figs:cylinder_Cd}. 
Following the initial transient phase, 
the flow settles into a stable periodic regime where $C_L$ oscillates symmetrically around zero, 
and $C_D$ fluctuates around a mean value. 
The convergence of these results is verified across different particle resolutions. 
Table \ref{tab:validation} provides a quantitative comparison 
of the mean drag coefficient, 
lift amplitude, and Strouhal number against established benchmarks. 
At the finest resolution ($dp=0.025$), 
we obtained a mean $C_D$ of $1.59$ and an $St$ of $0.179$. 
While slightly higher than values from grid-based Eulerian methods 
(e.g., Brehm et al. \cite{brehm2015locally}), 
our results show excellent agreement with recent Lagrangian SPH studies, 
such as those by Zhang et al. \cite{zhang2023lagrangian}. 
It is noteworthy that considerable scatter in force coefficients 
has been extensively documented in the literature for flows 
at moderate Reynolds numbers, 
suggesting a high sensitivity to numerical configurations 
or the potential existence of multiple realizable solutions \cite{zhang2023lagrangian}. 
Consequently, the deviations observed in the present study are not unexpected.
\subsection{Block sliding}\label{sec:Sliding}
\begin{figure}[htb!]
	\centering
	\includegraphics[width=0.6\textwidth] {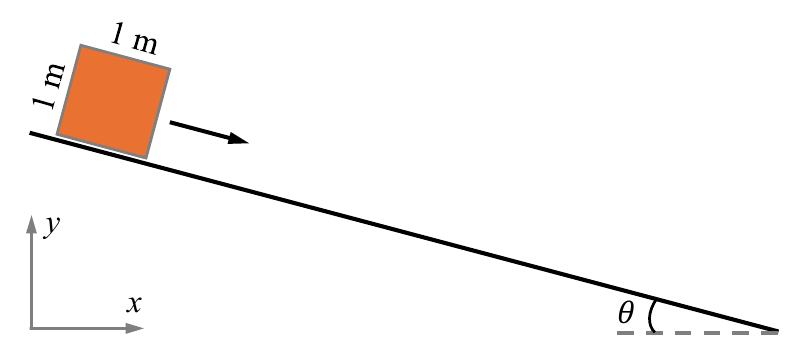}
	\caption{Block sliding: Model setup.}
	\label{figs:sliding_setup}
\end{figure}

To validate the accuracy of the proposed contact algorithm, 
specifically between an elastic solid and a shell structure, 
the block sliding benchmark is conducted \cite{zhu2022dynamic}. 
The geometric configuration of the problem 
is depicted in Fig.~\ref{figs:sliding_setup}. 
An elastic cube with a side length of $1 \, \text{m}$ 
is positioned on an inclined plane with a tilt angle of $\theta = 10^{\circ}$. 
A distinguishing feature of this test, 
compared to standard rigid-wall boundary treatments 
(e.g., in Ref.~\cite{zhu2022dynamic}), 
is that the fixed wall is modeled here using a single-layer shell formulation. 
This setup rigorously tests the capability of the contact forcing 
to prevent penetration while maintaining the correct sliding mechanics.

\begin{figure}[htb!]
	\centering
	\includegraphics[width=1.0\textwidth] {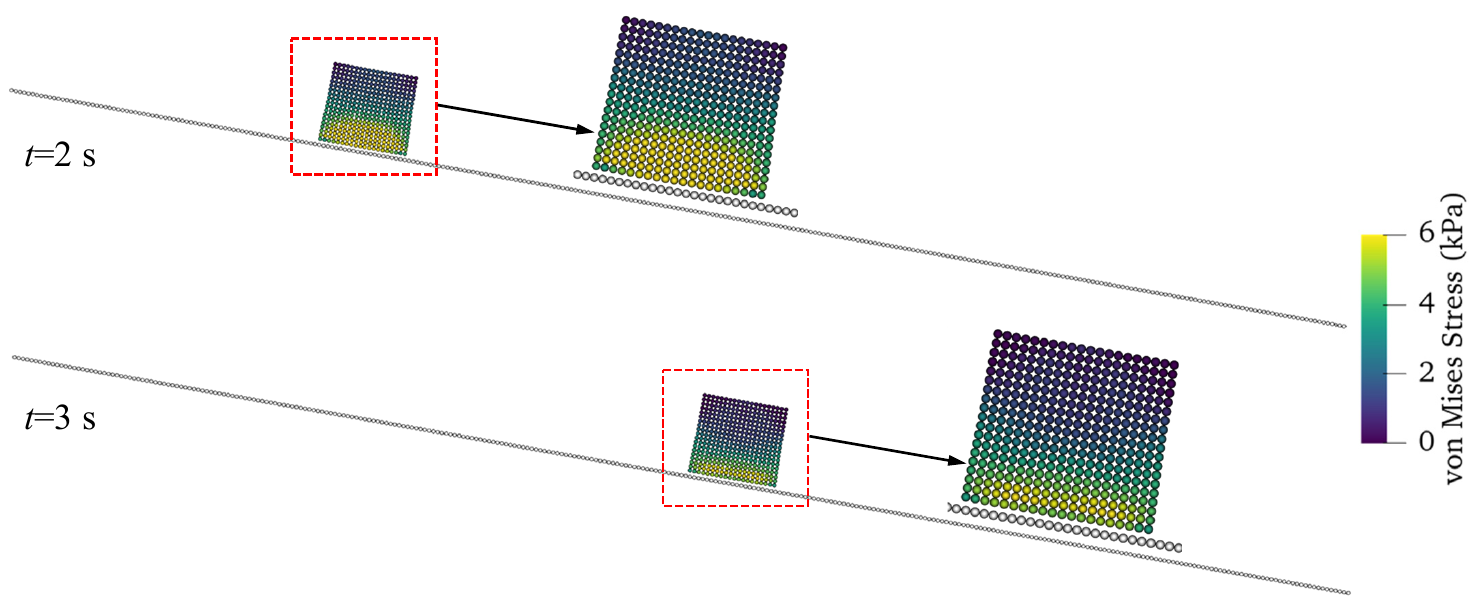}
	\caption{Block sliding: SPH simulation snapshots at $t=2 \, \text{s}$ and $t=3 \, \text{s}$.}
	\label{figs:sliding_snapshot}
\end{figure}

\begin{figure}[htb!]
	\centering
	\includegraphics[width=0.6\textwidth] {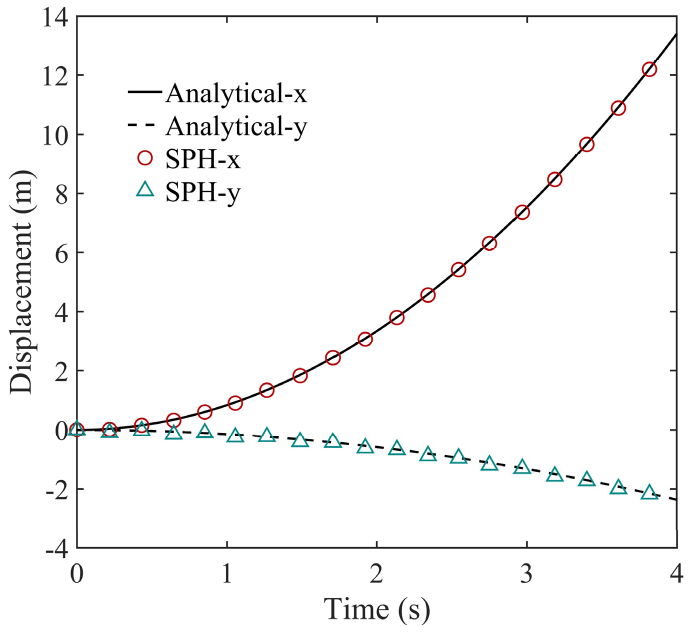}
	\caption{Block sliding: Time history of the block center displacement in the $x$- and $y$-directions.}
	\label{figs:sliding_disp}
\end{figure}

The material properties of the elastic block are set as follows: 
density $\rho = 1000 \, \text{kg/m}^3$, 
Young's modulus $E = 1.0 \times 10^5 \, \text{Pa}$, 
and Poisson's ratio $\nu = 0.45$. 
The system is driven solely by a gravitational acceleration of $g = 9.8 \, \text{m/s}^2$. 
The initial particle spacing is set as $dp =0.05 \ \text{m}$.
Assuming a frictionless interface between the elastic block and the shell boundary, 
the theoretical trajectory of the block center of mass
can be derived from Newton's laws of motion:
\begin{equation}
    \begin{aligned}
        & x(t) = x_0 + \frac{1}{2} g \sin\theta \cos\theta \, t^2, \\
        & y(t) = y_0 - \frac{1}{2} g \sin^2\theta \, t^2,
    \end{aligned}
    \label{eq:sliding_analytical}
\end{equation}
where $x_0$ and $y_0$ represent the initial coordinates.

Fig.~\ref{figs:sliding_snapshot} presents the simulation snapshots at $t=2 \, \text{s}$ and $t=3 \, \text{s}$. 
Qualitatively, 
the elastic block maintains its geometric integrity throughout the sliding process. 
Crucially, 
the interaction between the solid particles 
and the underlying shell particles remains stable, 
with no unphysical penetration or separation observed at the interface.

For a quantitative assessment, 
the time evolution of the block's displacement is analyzed. 
Fig.~\ref{figs:sliding_disp} compares the numerical results 
against the analytical solution given in Eq.~\eqref{eq:sliding_analytical}. 
The SPH-predicted displacements 
in both the horizontal ($x$) and vertical ($y$) directions 
exhibit excellent agreement with the theoretical curves. 
These results confirm that the proposed method 
effectively captures the dynamic contact behavior 
between elastic bodies and shell-based boundaries.
\subsection{Three rings contact}\label{sec:three_rings}
\begin{figure}[h]
	\centering
	\includegraphics[width=0.5\textwidth]{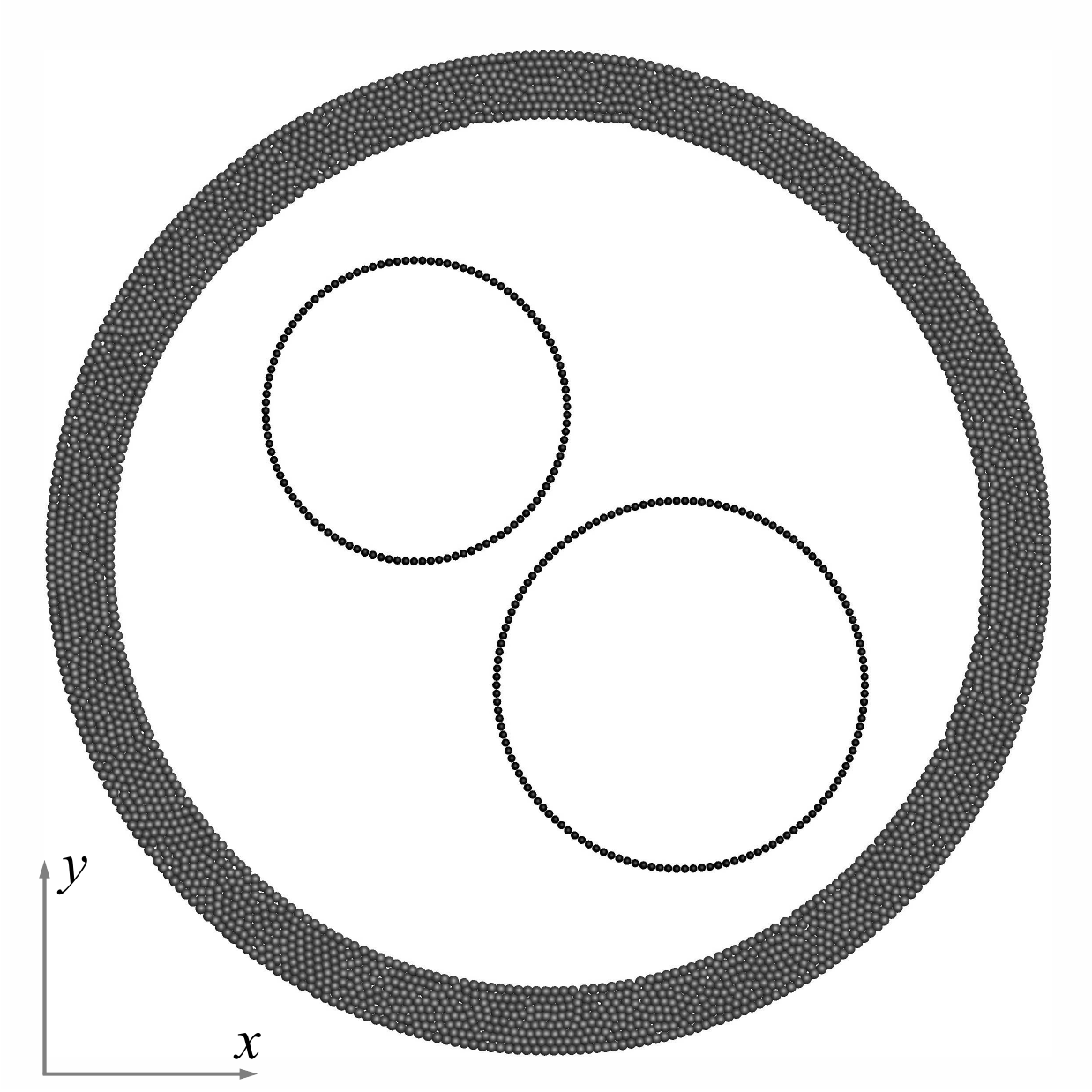}
	\caption{Three rings contact: Initial configuration.}
	\label{fig:ring_configuration}
\end{figure}

The solid-shell, shell-shell and shell-self contacts of three rings
is designed to further evaluate the performance of the proposed framework 
following Ref. \cite{yang2008large}. 
The initial configuration for this problem 
is shown in Figure \ref{fig:ring_configuration}. 
The dimensions of the rings are as follows:  
the inner and outer diameters of the largest ring are 26 and 30, respectively. 
For the medium-sized ring, these diameters are 10 and 12, 
while the smallest ring has inner and outer diameters of 8 and 10. 
The center of the largest ring is located at \((0, 0)^T\), 
the center of the medium-sized ring is at \((3.95, -3.95)^T\), 
and the smallest ring is at \((-3.95, 4.25)^T\). 
The three rings are modeled 
with Young's moduli \(E_1 = 10,000\) for the smallest ring, 
\(E_2 = 2250\) for the medium ring, 
and \(E_3 = 288,000\) for the largest ring. 
The Poisson's ratio is \(\nu = 0.125\) for all rings, 
and the initial densities are \(\rho_1 = 0.1\), \(\rho_2 = 0.01\), 
and \(\rho_3 = 1.0\) for the smallest, medium, and largest rings, respectively.
\begin{figure}[h]
	\centering
	\includegraphics[width=0.9\textwidth]{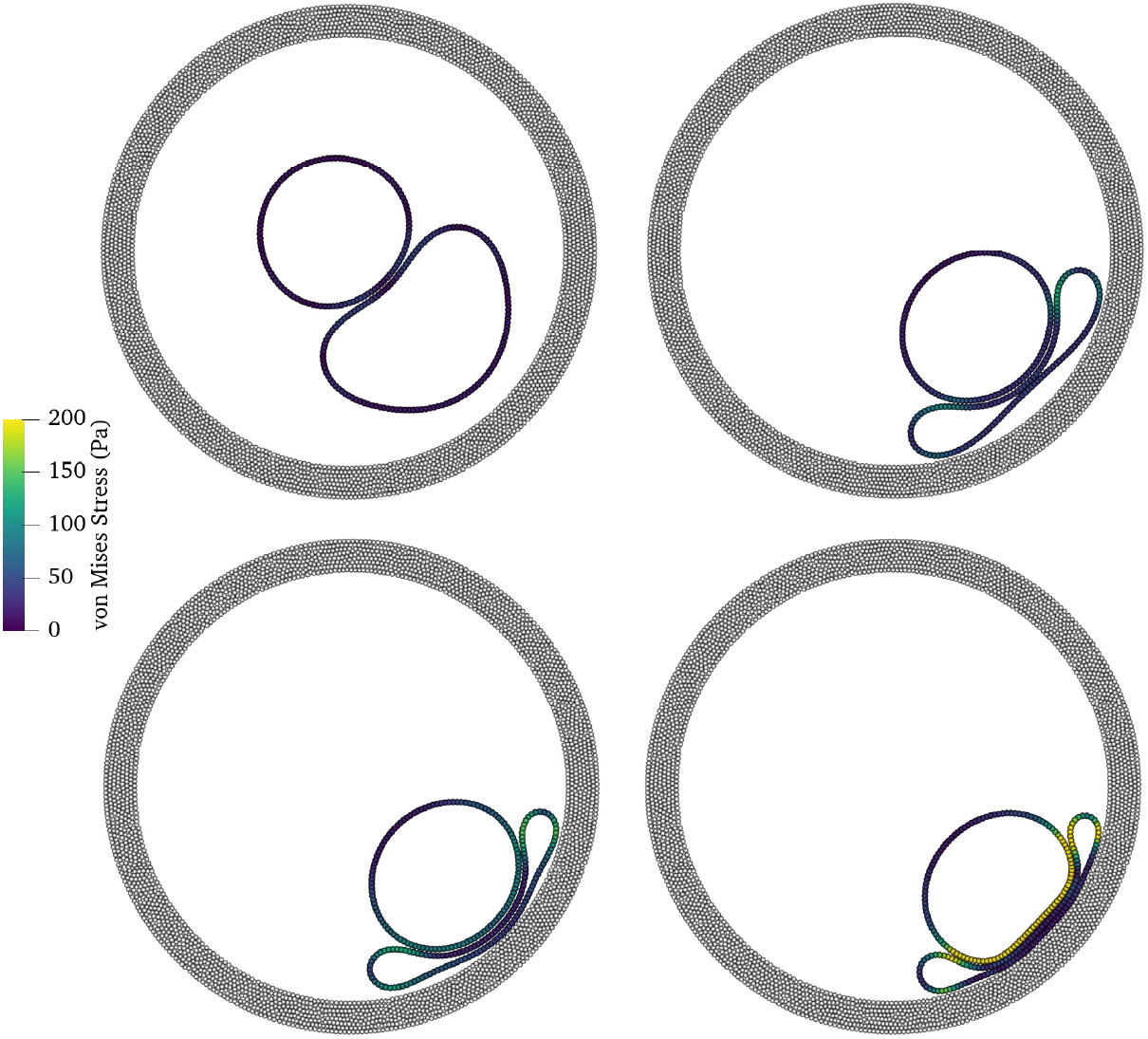}
	\caption{Three rings contact: Deformed configurations colored by von Mises stress at several temporal instants.}
	\label{fig:ring_deform}
\end{figure}

The external boundary of the largest ring is fixed, 
while the smallest ring, 
positioned inside the largest and adjacent to a third medium-sized ring, 
is initially given a velocity of $\bm{v} = (30.0, -30.0)^T$. 
As the smallest ring collides with the medium-sized ring, 
both rings move together toward the inner surface of the largest ring. 
During this process, self-contact occurs within the medium-sized ring, 
as illustrated in the deformed configurations in Figure \ref{fig:ring_deform}. 
As shown in the results, 
the framework efficiently handles the large deformations 
and contact that occur during the impact, 
accurately identifying self-contact regions, 
capturing the complex dynamics of the interacting rings, 
and aligning well with the findings presented in Ref. \cite{yang2008large} 
(refer to Figure 16 in their work). 
\subsection{Oil tank collision}\label{sec:oil_tank}
\begin{figure}[h]
	\centering
	\includegraphics[width=\textwidth]{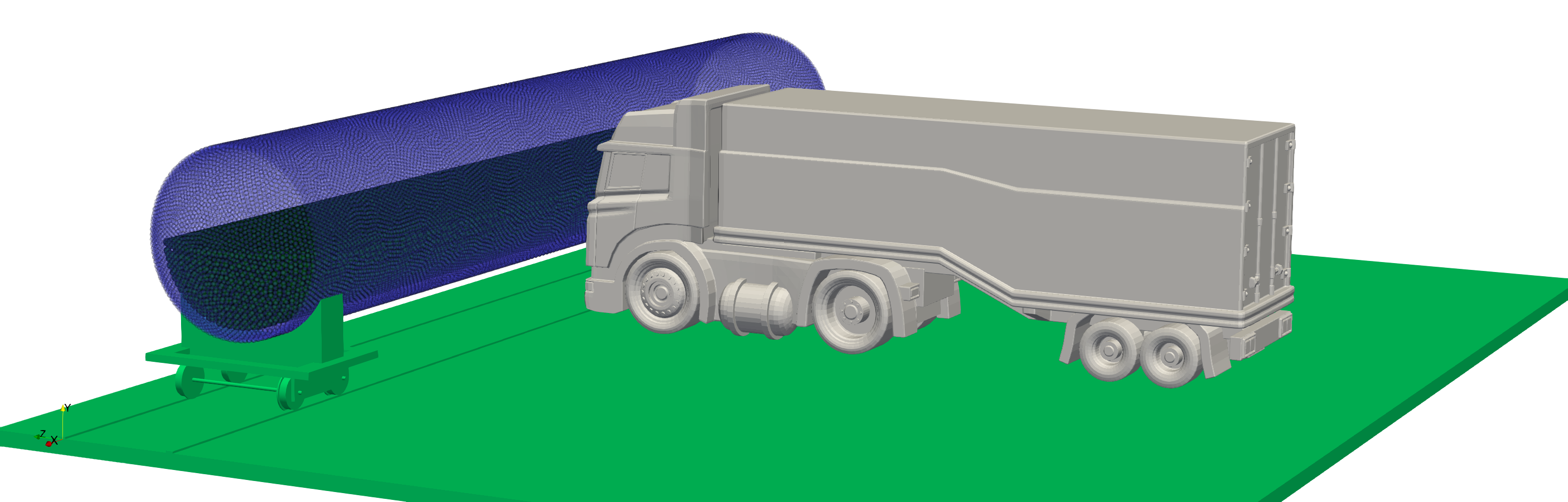}
	\caption{Oil tank collision: Initial configuration.}
	\label{fig:oil_tank_configuration}
\end{figure}

\begin{figure}[htb!]
	\centering
	\includegraphics[width=0.63\textwidth]{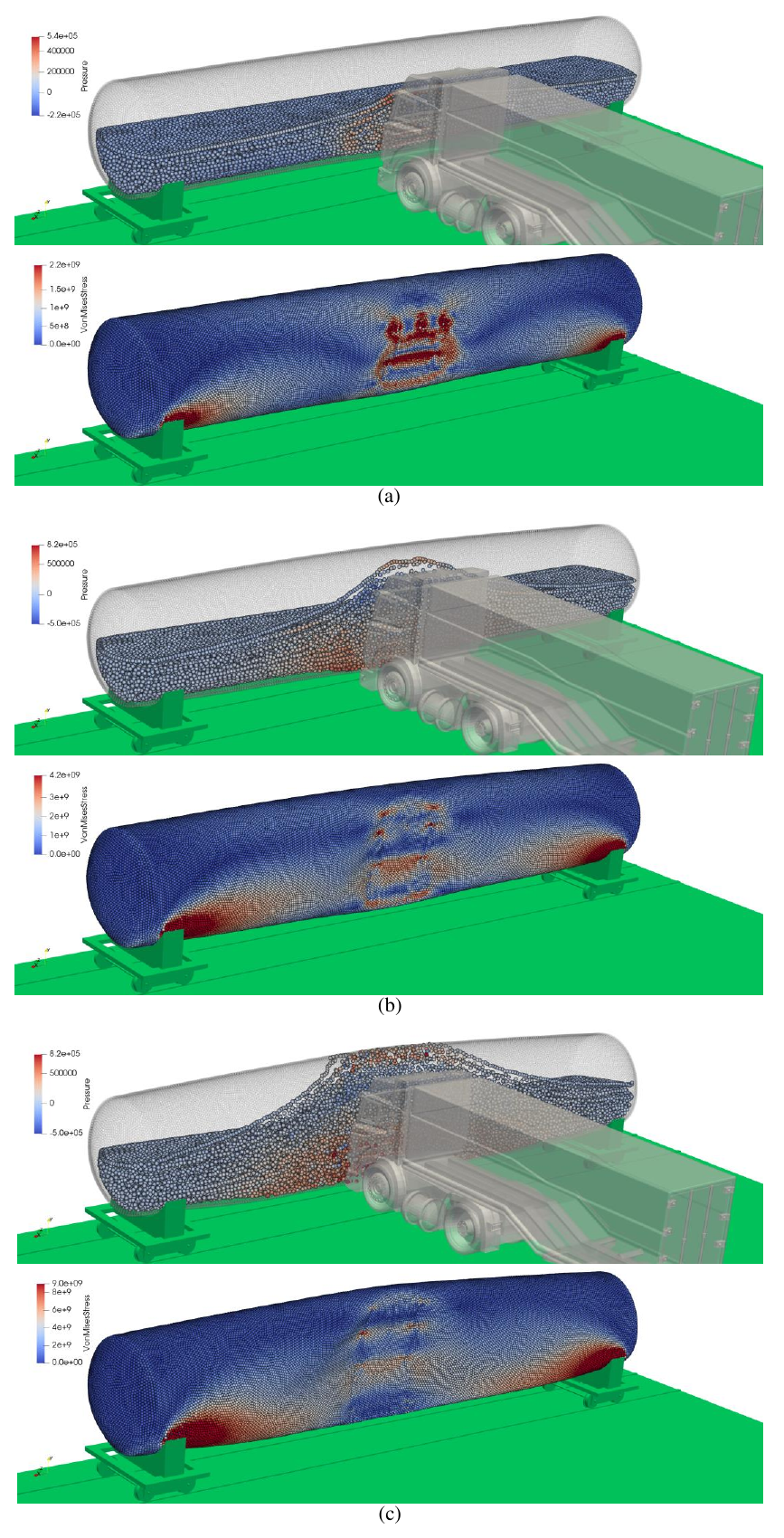}
	\caption{Oil tank collision: Deformed configurations colored by pressure and von Mises stress at several temporal instants.}
	\label{fig:oil_tank_pressure_stress}
\end{figure}

In this section, 
we present a collision simulation of a oil tank car 
impacted by a truck at a level crossing following Ref. \cite{tian2020collision}, 
to analyze the stress distribution 
and displacement behavior of the tank car 
during the collision. 
As shown in Figure \ref{fig:oil_tank_configuration}, 
the oil tank car is treated by shell model
with a thickness of $d = 0.011 \, \text{m}$, 
a density of $\rho_0^S = 7850.0 \, \text{kg/m}^3$, 
an elastic modulus of $E = 206.0 \, \text{GPa}$, 
and a Poisson's ratio of $\nu = 0.28$. 
The tank car is half-filled by oil, 
whose material properties are: 
a density of $\rho_0^F = 700.0 \, \text{kg/m}^3$, 
a dynamic viscosity of $\mu =0.1 \, \text{Pa·s}$. 
The truck is moving at 20 m/s, 
colliding with the middle of the tank. 
As shown in Figure \ref{fig:oil_tank_pressure_stress}, 
the deformation of tank is well captured, 
and the stress of tank and the pressure of oil are smooth.

In this section, 
we present a collision simulation of an oil tank car
impacted by a truck at a level crossing, 
following Ref. \cite{tian2020collision}, 
to analyze the stress distribution and displacement behavior of the tank car 
during the collision. 
As depicted in Figure \ref{fig:oil_tank_configuration}, 
the oil tank car is modeled using a layer of shell particles 
with a thickness of $d = 0.011 , \text{m}$, 
a density of $\rho_0^S = 7850 , \text{kg/m}^3$, 
an elastic modulus of $E = 206 , \text{GPa}$, 
and a Poisson's ratio of $\nu = 0.28$. 
The tank is half-filled with oil, 
characterized by a density of $\rho_0^F = 700 , \text{kg/m}^3$ 
and a dynamic viscosity of $\mu = 0.1 , \text{Pa·s}$. 
The truck, traveling at 20 m/s, 
collides with the center of the tank. 
As shown in Fig.~\ref{fig:oil_tank_pressure_stress}, 
the proposed framework accurately captures the tank deformation 
at several representative time instants. 
Despite the strongly coupled and highly dynamic setting 
involving large deformation, fluid motion, and contact events, 
the stress field in the tank and the oil pressure remain smooth 
and free of spurious oscillations. 
These results demonstrate the capability of the proposed framework 
to handle complex interaction scenarios in an industrially relevant setting.

\section{Concluding remarks}\label{sec:conclusion}
In this study, 
we have proposed a unified Smoothed Particle Hydrodynamics (SPH) framework 
for simulating shell-related interactions, 
including one-sided fluid-shell, solid-shell, shell-shell, and shell-self contact dynamics. 
A key contribution of this work is 
the introduction of imaginary shell contact particles, 
projected from real shell particles along the normal direction, 
which effectively transforms the reduced-dimensional shell model 
into a full-dimensional one. 
This projection method not only ensures kernel completeness in fluid-shell interactions 
but also extends the developed solid contact models 
to accurately handle complex shell-related contact scenarios.
The developed framework has been validated through a series of benchmark tests. and the results show that 
the proposed method can effectively simulate challenging scenarios 
involving shell structures and multiple types of interactions
with reliable performance. 

While the proposed framework provides a robust foundation 
for shell-related interaction modeling, 
further extensions are possible. 
A natural next step is to extend the formulation 
to support two-sided fluid–shell interactions, 
in which the shell is coupled to fluids on both sides.
%
%
\section*{CRediT authorship contribution statement}
{\bfseries  D. Wu:} Conceptualization, Methodology, Investigation, Visualization, Validation, Formal analysis, Writing - original draft, Writing - review and editing; 
{\bfseries  S.H. Zhang:} Visualization, Validation, Writing - original draft, Writing - review and editing; 
{\bfseries  W.Y. Kong:} Methodology, Investigation, Validation;
{\bfseries  X.Y. Hu:} Supervision, Methodology, Investigation, Writing - review and editing.
%
%
\section*{Declaration of competing interest }
The authors declare that they have no known competing financial interests 
or personal relationships that could have appeared to influence the work reported in this paper.
%
%
\section*{Acknowledgments}
D. Wu and X.Y. Hu would like to express their gratitude to the German Research Foundation (DFG) 
for their sponsorship of this research under grant number DFG HU1527/12-4.
\clearpage
%
%
\section*{References}
\vspace{-0.8cm}
\renewcommand{\refname}{}
\bibliographystyle{elsarticle-num}
\bibliography{IEEEabrv,mybibfile}
\end{document}